\documentclass[fleqn,usenatbib]{mnras}

\usepackage{newtxtext,newtxmath}

\usepackage[T1]{fontenc}

\DeclareRobustCommand{\VAN}[3]{#2}
\let\VANthebibliography\thebibliography
\def\thebibliography{\DeclareRobustCommand{\VAN}[3]{##3}\VANthebibliography}

\usepackage{graphicx}	
\usepackage{amsmath}	
\usepackage{anyfontsize}

 \usepackage[flushleft]{threeparttable}
\def\msun{M$_{\odot}$}

\def\nh{$N_{\rm H}$}

\def\degs{\ifmmode ^{\circ}\else$^{\circ}$\fi}
\def\amin{\ifmmode ^{\prime}\else$^{\prime}$\fi}
\def\asec{\ifmmode ^{\prime\prime}\else$^{\prime\prime}$\fi}
   
\def\degs{\ifmmode ^{\circ}\else$^{\circ}$\fi}
\def\amin{\ifmmode ^{\prime}\else$^{\prime}$\fi}
\def\asec{\ifmmode ^{\prime\prime}\else$^{\prime\prime}$\fi}
\def\fss{\hbox{$.\!\!^{\rm s}$}}        
\def\farcs{\hbox{$.\!\!^{\prime\prime}$}}  
\def\h{$^{\rm h}$}
\def\m{$^{\rm m}$}

\newcommand{\flux}{erg~s$^{-1}$~cm$^{-2}$}
\def\src{J0720}
\def\xmm{\textit{XMM-Newton}}

\def\eros{\textit{eROSITA}}
\def\gaia{\textit{Gaia}}
\def\ps{Pan-STARRS}
\def\pb{$P_{\rm orb}$}
\usepackage{ulem}
\usepackage{multirow}

\title[The polar MASTER OT J072007.30+451611.6]{MASTER OT
J072007.30+451611.6: A Polar with Strong Optical Variability and Suppressed 
He~II Emission}

\author[Bobakov et al.]{
A. V. Bobakov,$^{1}$\thanks{E-mail: bobakovalex@gmail.com}
S. V. Zharikov,$^{2}$
A. V. Karpova,$^{1}$
D. A. Zyuzin,$^{1}$ 
A. Yu. Kirichenko,$^{2,1}$
Yu. A. Shibanov,$^{1}$
\newauthor
\ R. Karimov,$^{3}$  
N. L. Vaidman,$^{4,5}$
Sh. T. Nurmakhametova,$^{4}$
M. R. Gilfanov,$^{6,7}$
and R. Michel$^{2}$
\\
$^1$Ioffe Institute, 26 Politekhnicheskaya, St. Petersburg, 194021,  Russia \\
$^2$Instituto de Astronom\'ia, Universidad Nacional Aut\'onoma de M\'exico, Apdo. Postal 106, Baja California, M\'exico, 22860   \\
$^3$Ulugh Beg Astronomical Institute, Uzbekistan Academy of Sciences, Tashkent, 100052, Uzbekistan \\
$^{4}$Faculty of Physics and Technology, Al-Farabi Kazakh National University, Al-Farabi Ave., 71, 050040, Almaty, Kazakhstan\\
$^{5}$Fesenkov Astrophysical Institute, Observatory, 23, Almaty, 050020, Kazakhstan\\
$^6$Space Research Institute of the Russian Academy of Sciences, Profsoyuznaya Str. 84/32, 117997 Moscow, Russia \\ 
$^7$Max-Planck-Institut für Astrophysik, Karl-Schwarzschild-Str. 1, D-85741 Garching, Germany
}

\date{Accepted XXX. Received YYY; in original form ZZZ}

\pubyear{2025}

\begin{document}
\label{firstpage}
\pagerange{\pageref{firstpage}--\pageref{lastpage}}
\maketitle

\begin{abstract}
The transient optical source MASTER OT J072007.30+451611.6 has been recently
discovered and proposed as a peculiar polar with an unusually high amplitude 
of the   orbital brightness variation in the optical of $\sim$3 mag. To clarify its nature, 
we performed multiband time-series optical photometry with 1.5-m class
telescopes and spectroscopy with the 10.4-m Gran Telescopio Canarias.    
We also analysed archival data of different optical surveys and detected 
the source in X-rays with the Spectrum-RG/\eros\ telescope. We confirm the
orbital period of $\approx$1.5 h with the high amplitude of the brightness
modulation. Compiling  survey  data, covering $\sim$19 yr, we find high 
and low brightness states of the object at time scales of years, likely explained by different accretion rates. Our data were obtained in the  
high brightness  state. Optical spectra with hydrogen and helium emission
lines, consisting of broad and narrow components, indicate the presence of
an accretion stream without disk. The Doppler tomography shows that the 
narrow component is mainly emitted from the Lagrangian L$_1$ point, while 
the broad component is from the region where the accretion stream interacts 
with the white dwarf magnetosphere. The ratio of equivalent widths of
\ion{He}{ii}~4686 and H$\beta$ emission lines is $<$0.4, which is curiously 
low for polars. The X-ray spectrum of the source can be described by the 
thermal plasma emission model with parameters consistent with values  
observed for polars.

\end{abstract}

\begin{keywords}
binaries: close -- novae, cataclysmic variables -- stars: individual: MASTER OT J072007.30+451611 
\end{keywords}



\section{Introduction}
 

MASTER OT J072007.30+451611.6 (hereafter \src) is an unusual non-eclipsing 
compact binary system recently discovered by \citet{Pogrosheva2018,denisenko2018}.
The optical data from the Catalina Sky Survey \citep[CSS;][]{drake} showed a large-amplitude ($\sim$3~mag) variability with the periodicity of 0.0627887~d 
\citep[1.50693~h;][]{denisenko2018}. The \src\ low-resolution optical 
spectrum was obtained in the Large Sky Area Multi-Object Fiber Spectroscopic 
Telescope (LAMOST, \citealt{Zhao_2012})
survey\footnote{\url{https://www.lamost.org/dr8/v2.0/spectrum/view?obsid=504804160}}.
It demonstrates relatively narrow Balmer emission lines which can  be a 
marker of an accretion process in the system. In addition, the source has an
X-ray counterpart\footnote{\url{https://www.aavso.org/vsx/index.php?view=detail.top&oid=621224}} designated as XMMSL2 J072007.4+451615 in the
\xmm\ Slew Survey Clean Source Catalog \citep[XMMSLEWCLN;][]{xmmsl}. Its  
observed flux in the 0.2--12 keV range is $(2.9 \pm 1.0)\times 10^{-12}$ \flux. 

\citet{denisenko2018}  classified \src\  as a polar or AM Herculis-type 
cataclysmic variable (CV). Polars represent a subclass of magnetic CVs (mCVs) in which the accreting white dwarf (WD) has a high magnetic field, 
$\sim$10--100 MG, which prevents the formation of an accretion disc 
\citep[e.g.][]{cropper1990}.  In this case, the accretion matter streams 
from the donor star directly along the magnetic field lines of the WD to 
its magnetic pole(s). We note that such strong optical orbital variability
of \src\ is not typical for polars. Usually, they show orbital variations 
of $\lesssim$2 mag. In addition, polars may switch between high and low accretion states, which strongly alters their brightness. 
\citet{denisenko2018} suggested that the highly modulated optical 
light curve of \src\ results from a large contribution of a hot spot at the WD 
surface.
%
%
%

There is another class of highly variable binaries with similar orbital periods, which consist of non-accreting millisecond pulsars and 
low-mass companions dubbed `spider pulsars' (e.g. \citealt{draghis2019, matasanchez2023,Bobakov2024}).
Their optical variability, typically of $\approx$2--4 mag, is a result of the
strong irradiation of the secondary by the pulsar wind. Optical spectra of some `spiders' demonstrate H and He emission lines, which can be attributed to the stellar wind and/or intrabinary shock \citep[e.g.][]{swihart2022}.

To investigate the nature of \src, we have performed its optical 
time-resolved photometric and spectroscopic observations. We also considered
optical data from different catalogues and X-ray data from the 
\eros\ telescope \citep{erosita2021} onboard the Spectrum–RG (SRG) 
observatory \citep{Sunyaev2021}. The catalogue data and our own 
observations are described in Sec.~\ref{sec:data} and analysed  in 
Sec.~\ref{sec:results}. We discuss and summarise the results in 
Sec.~\ref{sec:discussion}. 

\begin{table}
\centering
\caption{Log of the \src\ photometric observations.} 
\begin{threeparttable}
\centering
\label{tab:log-phot}
\begin{tabular}{cccl}
\hline
Date       & MJD  & Filter &  Exp. time \\
\hline
\multicolumn{4}{c}{\textbf{MAO}} \\
2023-11-12 & 60260 & $V$   & 120 s $\times$ 17 \\ 
           &       & $V$   & 180 s $\times$ 57 \\
           &       & $R$   & 180 s $\times$ 22 \\
           &       & $I$   & 180 s $\times$ 22 \\
\hline
\multicolumn{4}{c}{\textbf{OAN-SPM}} \\
2023-11-26 & 60274 & $V$   & 600 s $\times$ 11 \\
           &       &       & 300 s $\times$ 5 \\
           &       &       & 400 s $\times$ 1 \\
           &       & $R$   & 200 s $\times$ 6 \\
           &       &       & 300 s $\times$ 11 \\
           &       & $I$   & 200 s $\times$ 1 \\
           &       &       & 150 s $\times$ 2 \\
2023-12-12 & 60290 & $V$   & 90 s $\times$ 16 \\
           &       &       & 300 s $\times$ 1 \\
           &       & $R$   & 60 s $\times$ 17 \\
           &       &       & 100 s $\times$ 3 \\
2023-12-13 & 60291 & $V$   & 300 s $\times$ 27 \\
           &       & $R$   & 100 s $\times$ 28 \\
           &       & $I$   & 60 s $\times$ 28 \\            
2023-12-14 & 60292 & $V$   & 300 s $\times$ 19 \\
           &       & $R$   & 100 s $\times$ 19 \\
           &       & $I$   & 60 s $\times$ 19 \\
\hline
\multicolumn{4}{c}{\textbf{ATO}} \\
2025-01-10 & 60685 & $r$   & 20 s $\times$ 964 \\
2025-01-13 & 60688 & $i$   & 20 s $\times$ 402 \\
\hline
\end{tabular}
\begin{tablenotes}
\item \textit{Note.} MAO and OAN-SPM observations were performed with Johnson-Cousins filters and ATO -- with Sloan ones. 
\end{tablenotes}
\end{threeparttable}
\end{table}

\section{Observations and data reduction}  
\label{sec:data}


\subsection{Catalogue data}
\label{subsec:cat}
\begin{figure*}
\begin{minipage}[h]{1.\linewidth}
\center{\includegraphics[width=1.\linewidth,clip]{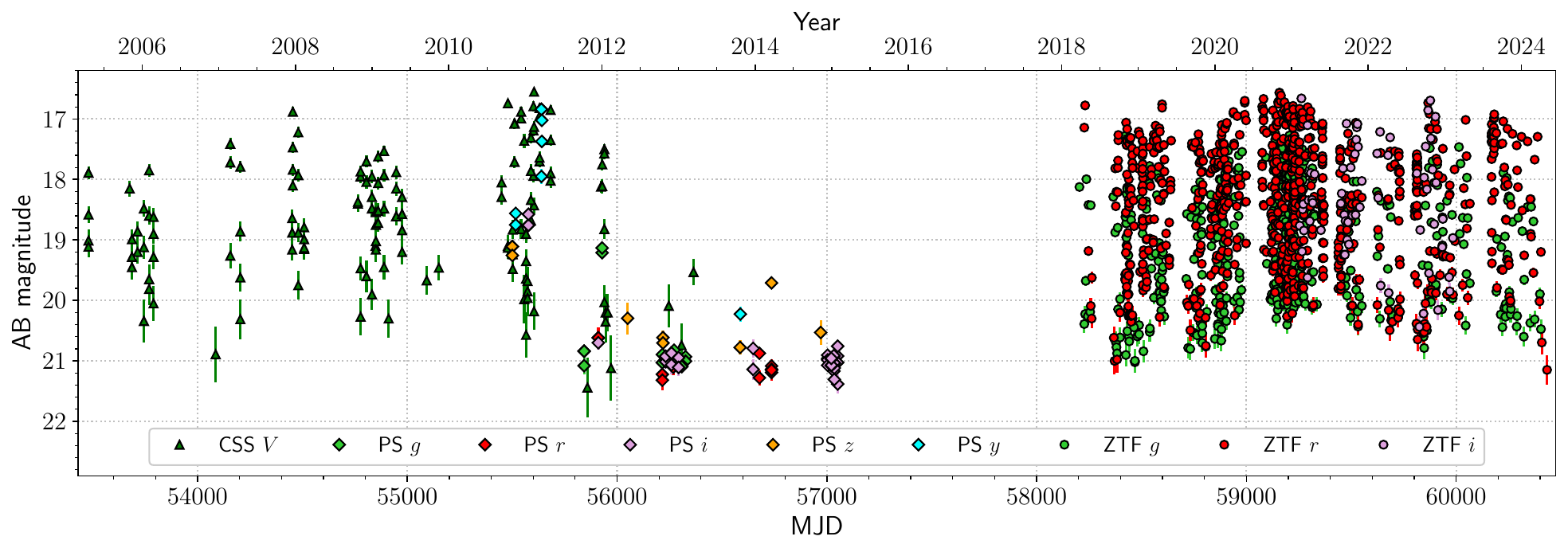}}
\end{minipage}
\caption{ J0720 light curves obtained from the CSS, \ps\ and ZTF data,
demonstrating changes between high and low brightness states. Data from 
different surveys and bands are shown by various symbols as indicated in 
the legend  (PS = \ps).}
\label{fig:panztf}
\end{figure*}

We found that \src\ is present in different optical catalogues.
As noted in \citet{denisenko2018}, it was identified as CSS J072007.4+451615
in the CSS catalogue. In addition, in the \gaia\ DR~3 catalogue \citep{gaia2016,gaia-dr3-2023} \src\ is designated as 974157681483861248, and has the coordinates R.A. = 07\h20\m07\fss3812(5), 
Dec. = +45\degs16\amin11\farcs507(1) and the mean magnitude $G=20.9$.
No parallax measurements were reported for the source. 
It is also identified in the Panoramic Survey Telescope and Rapid Response System (\ps) DR~2 catalogue \citep{ps2020}, 
the AllWISE Source Catalog (WISEA,  \citealt{allwise}) 
and the Zwicky Transient Facility (ZTF, \citealt{ztf}) DR~22 archive data.
The \ps\ and ZTF data confirm the strong variability of \src\, previously identified in the CSS data (Fig.~\ref{fig:panztf}).
%
%
%
%
%
%
\begin{figure}
\begin{minipage}[h]{1.\linewidth}
\center{\includegraphics[width=1.\linewidth,clip]{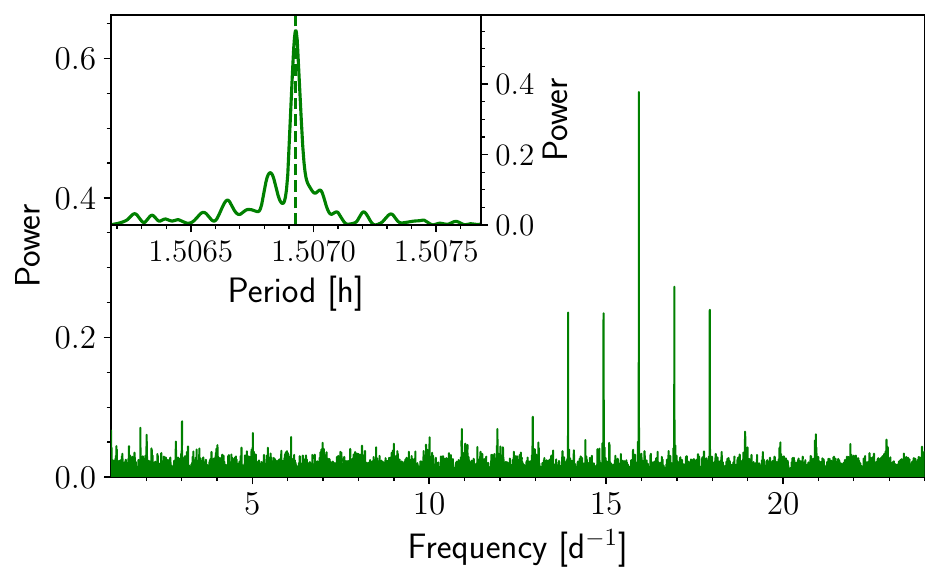}}
\end{minipage}
\caption{Lomb-Scargle periodogram calculated for the ZTF data in the $r$ band.
The best period corresponding to the highest peak, enlarged in the inset, is indicated by the dashed line. 
}
\label{fig:period-ztf}
\end{figure}
%
%
\subsection{Optical photometry and spectroscopy}
\label{subsec:phot-spec}
The time-series photometry of \src\ was performed using 1.5-m telescopes at
the Maidanak Astronomical Observatory (MAO) in Uzbekistan, Assy-Turgen 
observatory (ATO) in Kazakhstan, and Observatorio Astron\'omico Nacional San
Pedro M\'artir (OAN-SPM) in Mexico. 
Standard data reduction and photometric calibrations were applied using a
set of \ps\ field stars around the target and corresponding {\sc iraf} 
tasks. The log of observations is given in Table~\ref{tab:log-phot}.

The time-series spectroscopy was performed with the 10.4-m Gran 
Telescopio Canarias (GTC). The first run of long-slit observations was carried out in February 2024 through the Director’s Discretionary Time (DDT)\footnote{Program GTC11-23BDDT, PI A. Kirichenko.} with the upgraded Optical System for Imaging and low-intermediate Resolution Integrated Spectroscopy (OSIRIS+) instrument.  We obtained four consecutive
spectra with an exposure time of 400 s using the R1000B grism through the 0.8
arcsec slit. The resulting spectral resolution was 5.4~\AA. 
We obtained an additional set of observations in November 2024\footnote{Program GTC1-24BMEX, PI S. Zharikov.} using the same instrument and the R2000B and R2500R grisms. The full binary
orbit was covered with ten consecutive exposures for each grism. 
The spectral resolutions were 4.4~\AA\ (R2000B) and 3.4~\AA\ (R2500R).
The log of observations is given in Table~\ref{tab:log-spec}.

The data reduction was performed with the semi-automated pipeline packages
{\sc pypeit} \citep{pypeit:zenodo,pypeit:joss_pub} and {\sc iraf}. 
Flux calibration was achieved using the Feige~110 spectrophotometric standard.
%
%
\begin{table}
\centering
\caption{Log of the \src\ spectroscopic observations with the GTC.} 
\label{tab:log-spec}
\begin{tabular}{ccccl}
\hline
Date       & MJD   & Grism  & Range, \AA & Exp. time \\  
\hline
2024-02-03 & 60343.1001  & R1000B & 3630--7000 & 400 s $\times$ 4 \\
2024-11-03 & 60617.1329  & R2000B & 3950--5700 & 535 s $\times$ 10 \\
           & 60617.1986  & R2500R & 5575--7685 & 530 s $\times$ 10 \\
\hline 
\end{tabular}
\end{table}
%
%
\subsection{X-ray data}
\label{subsec:xrays}
The \src\ field was observed by the SRG/\eros\ in the course of four all-sky surveys in 2020--2021 with the total vignetting corrected exposure time of $\approx 440$ s. \eros\ data were calibrated and processed
by the calibration pipeline at the Space Research Institute (IKI) based
on the \eros\ Science Analysis Software System (eSASS) \citep{Brunner2022}
and pre- and in-flight calibration data. 
\section{Data analysis and results}
\label{sec:results}
\subsection{Long- and short-term variability in the optical}
\label{subsec:lcs}
The long-term brightness variation of \src\ collected by different 
photometric sky surveys is shown in Fig.~\ref{fig:panztf}. It covers 
the time interval from 2005 to 2024. It can be seen that the source
demonstrates high (MJD~$\approx$~53400--56000 and 58200--60500) and
low (MJD~$\approx$~56000--57000) brightness states. In the high state,
it is strongly variable with $\Delta$mag~$\sim 3$--4 mag, apparently 
at time scales from several days to months. Transitions between high and low states are typical for mCVs \citep[e.g.][]{lathan1981}, while they are not observed for `spider' systems.

We applied the Lomb-Scargle periodogram \citep{lomb1976,scargle1982} 
analysis to search for periodicity in the $r$-band ZTF data, which spans
over six years and contains about 800 measurements. The resulting power
spectrum is shown in Fig.~\ref{fig:period-ztf}. The highest peak $f_{\rm ph}
= 15.92644(25)$~d$^{-1}$ corresponds to the period\footnote{The uncertainty 
is calculated as the half width at half maximum of the highest peak in the 
periodogram.} $P_{\rm ph}=1.506928(24)$~h. It is in agreement with
the value reported by \citet{denisenko2018} and proposed as the orbital 
period \pb\ of the system. The ZTF, ATO, MAO, and OAN-SPM data folded with
the obtained period are shown in Fig.~\ref{fig:ztf-LCs} (the zero phase 
corresponds to $t_0$~=~HJD 2460617.620472 which was selected from 
spectroscopy and Doppler tomography; see below).  All the light 
curves are consistent in shape and demonstrate strong asymmetry with
a fast increase of the brightness,  a plateau state at maximum brightness 
with small amplitude flickering during $\approx$0.3\pb\ followed by a more 
gradual decline in brightness until the minimum lasting $\approx$0.4\pb.
There is a hint of an intermediate minimum during brightness decline.
Comparison of the amplitude of orbital brightness variations with those in 
Fig.~\ref{fig:panztf} indicates that \src\ was in the high state during our 
photometric observations. 
\subsection{Optical spectra}
\label{subsec:opt-spec}

\begin{figure*}
\begin{minipage}[h]{1.\linewidth}
\center{
\includegraphics[width=0.48\linewidth,clip]{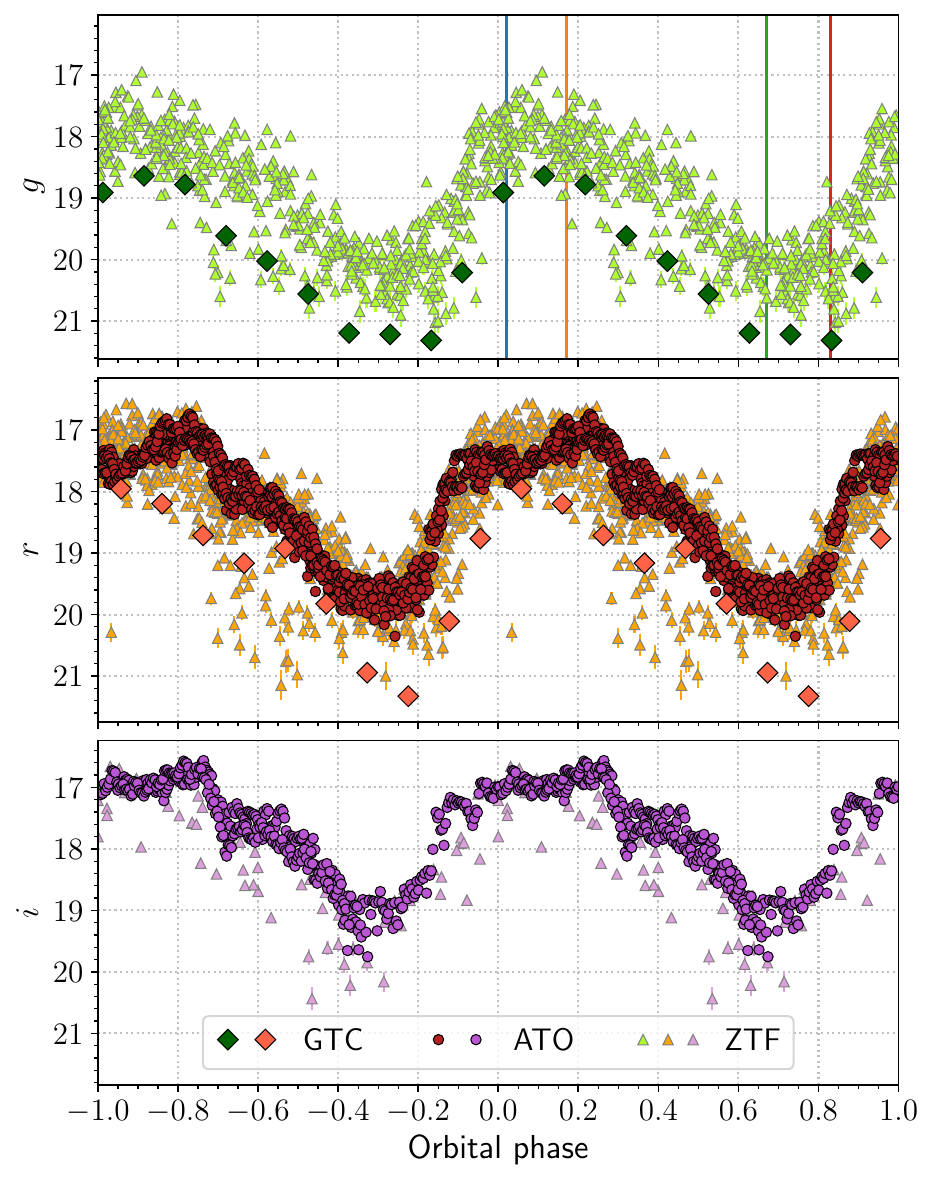}
\includegraphics[width=0.48\linewidth,clip]{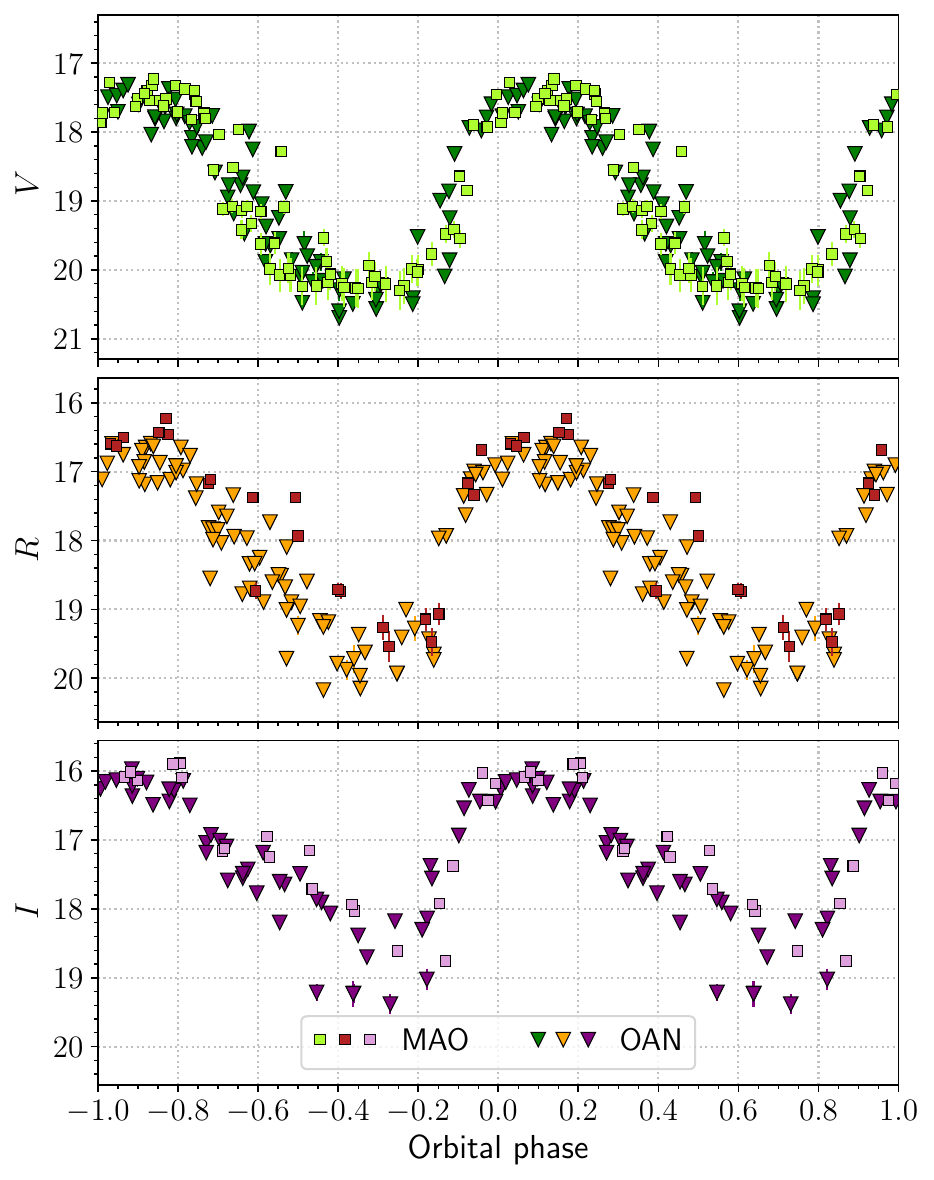}
}
\end{minipage}
\caption{Light curves of \src\ folded with the period of 1.506928 h. 
Data from different instruments are marked with different symbols as 
indicated in the legends. ZTF and ATO measurements, as well as magnitudes 
obtained from the GTC spectra in the AB system, are presented in the left 
panels, while MAO and OAN-SPM measurements in the Vega system -- in the 
right panels. Two periods are shown for clarity. The zero phase was 
selected based on spectroscopic data (see text below). Vertical colour 
lines in the top left panel mark the phases at which the spectra shown 
in Fig.~\ref{fig:opt-spec} were obtained.}
\label{fig:ztf-LCs}
\end{figure*}

\begin{figure*}
    \centering
     \includegraphics[width=\linewidth]{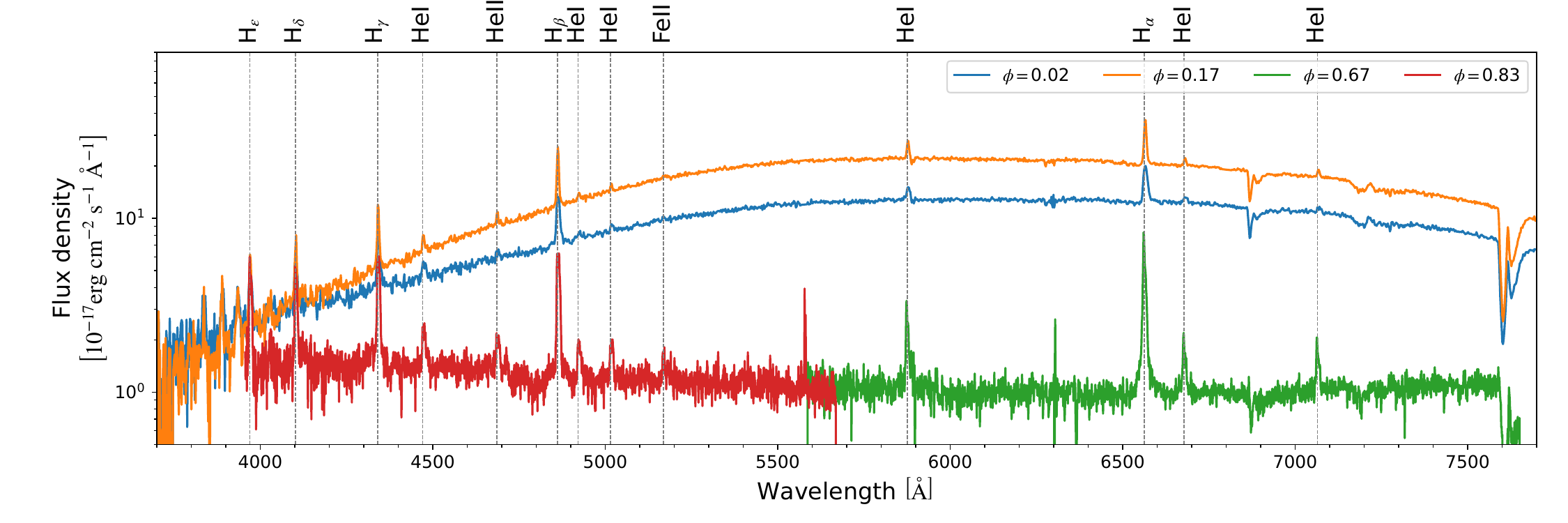} 
    \caption{Optical spectra of \src\ obtained at different orbital phases with the GTC using R1000B (orange and blue), R2000B (red) and R2500R (green) grisms.}
    \label{fig:opt-spec}
\end{figure*}

In Fig.~\ref{fig:opt-spec}, we show \src\ optical spectra obtained at 
orbital phases near the maximum and minimum of the object's brightness 
marked by vertical lines in the left-top panel of Fig.~\ref{fig:ztf-LCs}. 
The spectra demonstrate strong Balmer emission lines, which have been seen in the LAMOST data. 
In addition, weaker \ion{He}{I} and \ion{He}{II} emission 
lines are detected. The object shows substantial continuum variability 
with the highest amplitude in the 5000--7000 \AA\ range, that decreases 
to the bluer and redder parts of the spectrum and disappears below 4000 \AA.
To check the brightness state of \src\ during the spectroscopy runs, we 
extracted  magnitudes from the spectral data obtained in November 2024 
using the \ps\ $g$-filter transmission curve and a simple rectangular 
filter with the wavelength range of 5600--7400 \AA\ partially overlapping 
with the $r$ band. The results are presented in the top and middle 
subpanels of Fig.~\ref{fig:ztf-LCs}, left. The obtained light curves are 
compatible with the photometric ones, which shows that \src\ was in the 
high state during the spectral observations as well. The four spectra
obtained in February 2024 are consistent with the November ones at close 
orbital phases. 
\begin{figure*}
    \centering
     \includegraphics[width=\linewidth]{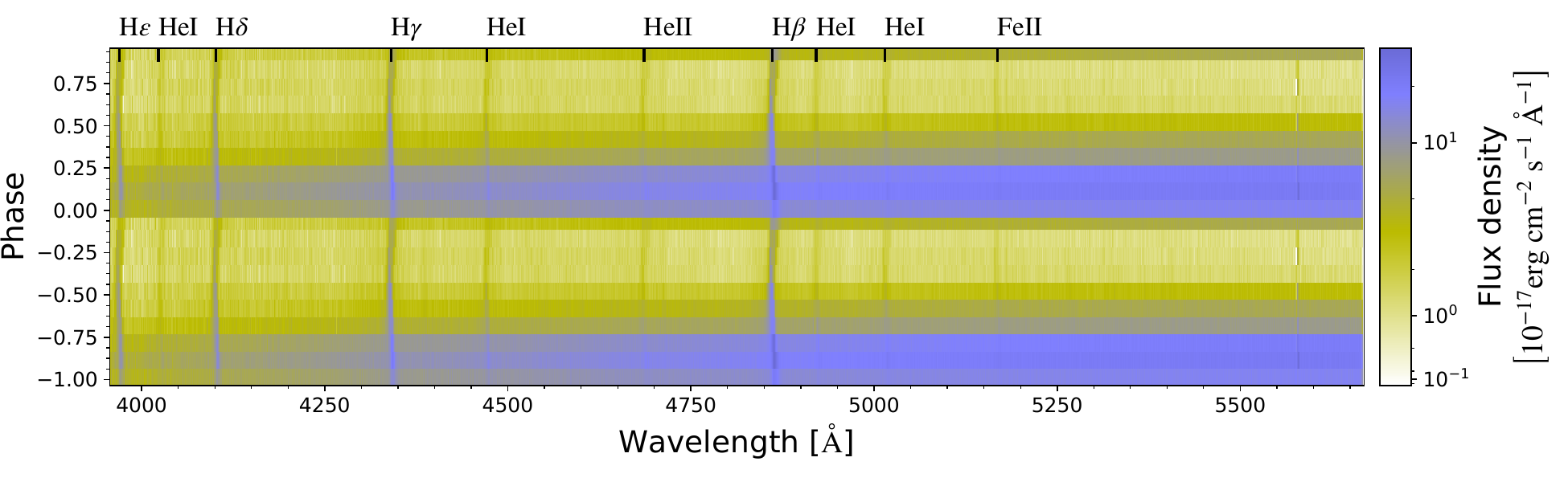} \\
    \includegraphics[width=\linewidth]{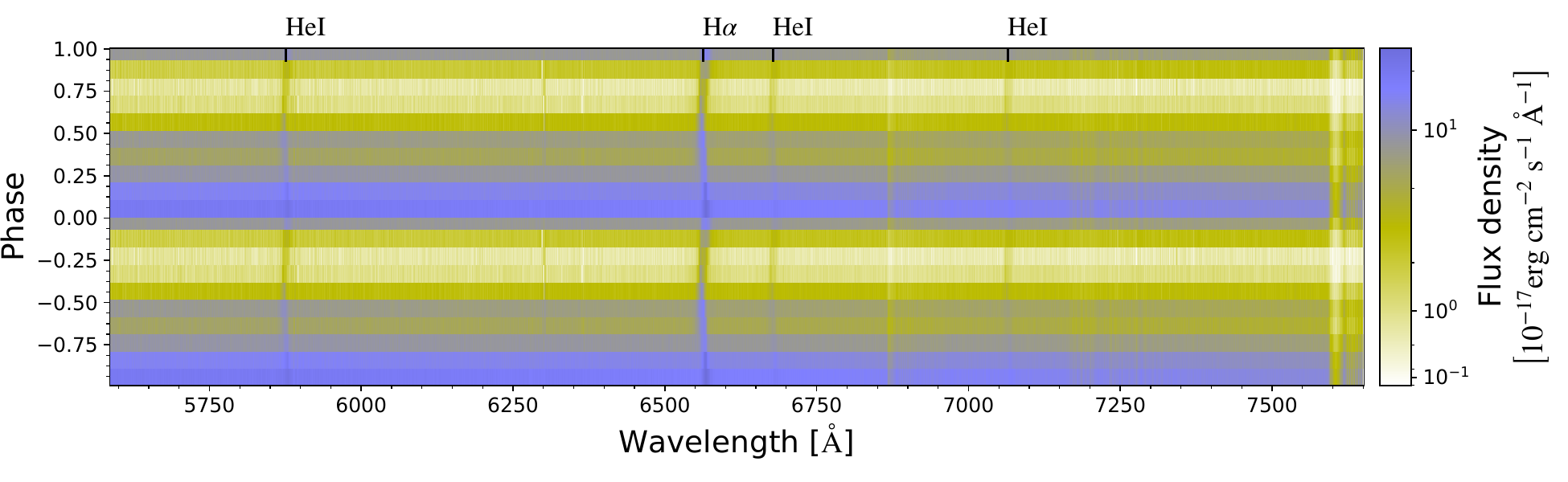}
    \caption{Trailed spectra of \src\ obtained with R2000B (top) and 
    R2500R (bottom) grisms. Two orbital cycles are presented for clarity.}
    \label{fig:trailed_full}
\end{figure*}
\begin{figure*}
\centering
\includegraphics[width=0.9\linewidth]{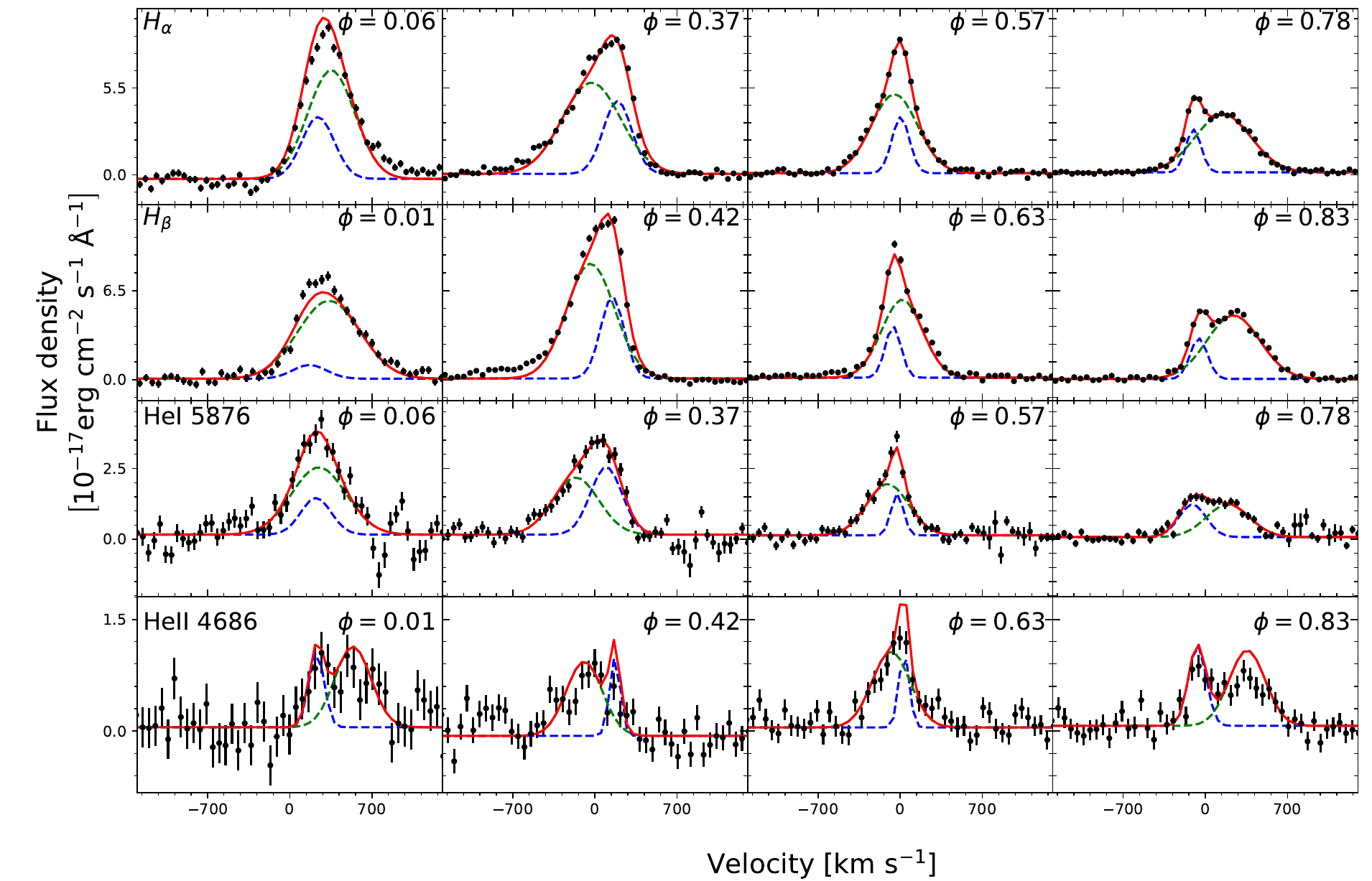}
\caption{Profiles of two Balmer lines (two upper rows) and He (two lower rows) spectral 
lines at different orbital phases $\phi$ indicated in the panels. 
The black points with error bars show continuum-subtracted data. 
In each panel, the red solid line is the best-fitting double-Gaussian 
model. The blue and green dashed lines show the best fit components of the model. }
\label{fig:lines_phase}
\end{figure*}
\begin{figure*}
\includegraphics[width=0.93\linewidth, bb = 0 60 1160 1000, clip=]{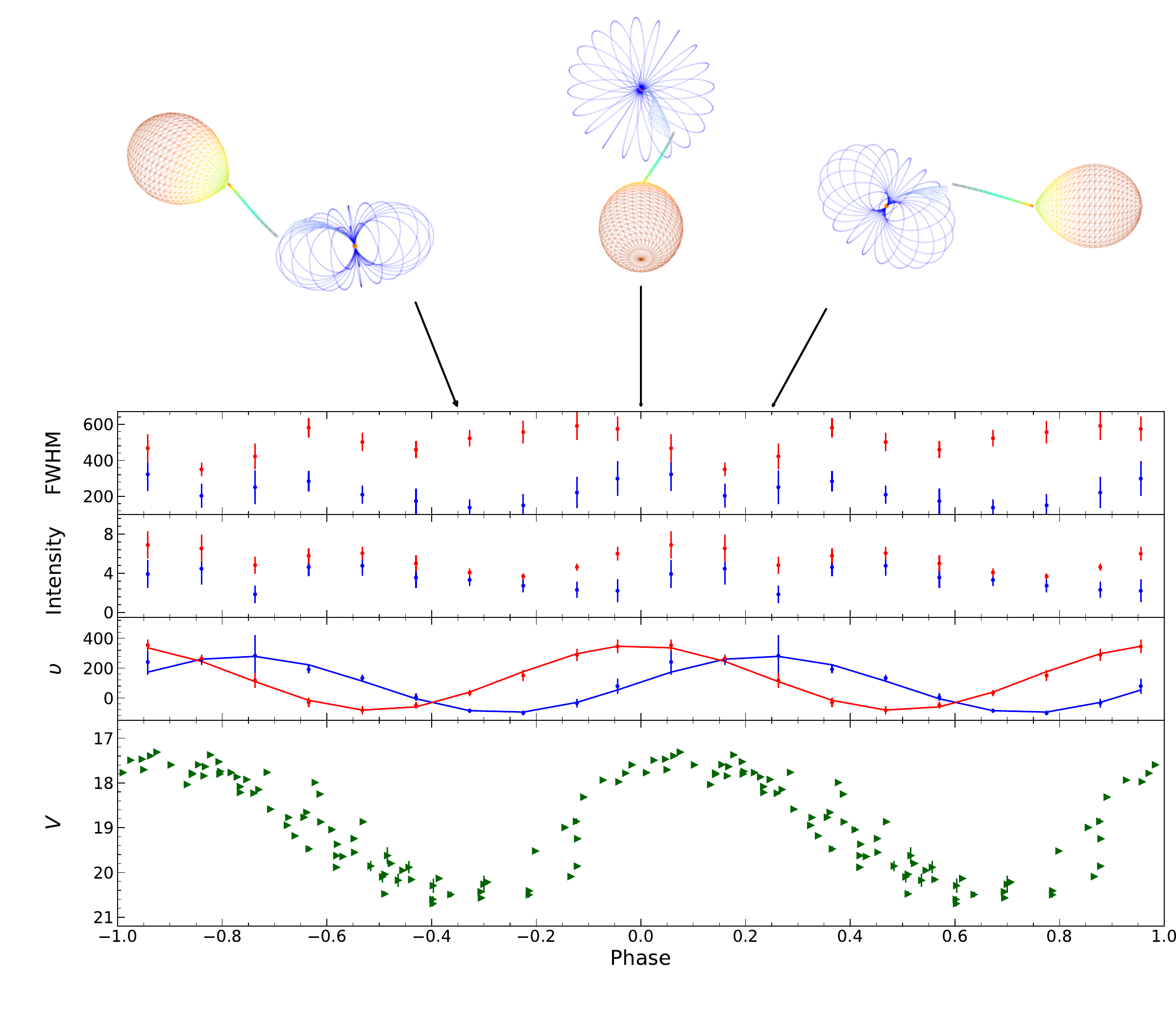}
\caption{Variations of the best-fitting parameters for the H$\alpha$ line 
profiles obtained with the double Gaussian model with the orbital phase.  
The top panel shows FWHMs in units of km~s$^{-1}$, the middle panel -- 
continuum-subtracted intensities in units of 10$^{-17}$~erg~cm$^{-2}$~s$^{-1}$~\AA$^{-1}$
, the third panel -- RVs in units of km~s$^{-1}$ and the bottom panel -- light curve in the $V$ band. 
Results for the broad and narrow line components are presented  
in red and blue, respectively. In the third panel, lines represent 
the sine fits of the RV curves. At the top of the plot the artistic views on the system configuration with parameters from Sect.~\ref{SysPar} in different orbital phases are shown. The size of the magnetosphere (with blue magnetic field lines), the accreting stream (green-blue), and the result of irradiation of the secondary (yellow) were selected arbitrary.}
\label{fig:lines_fit_pars}
\end{figure*}
%
%
%
\begin{figure*}
\centering
\includegraphics[width=0.8\linewidth]{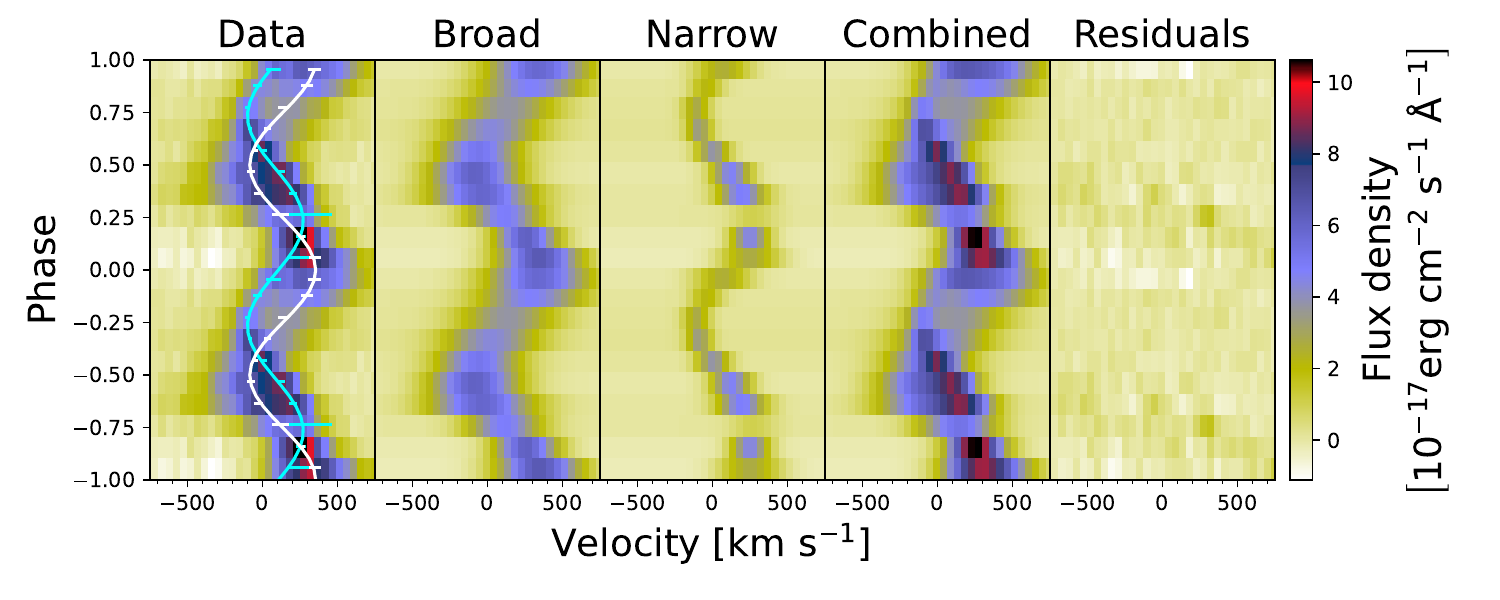} \\
\includegraphics[width=0.8\linewidth]{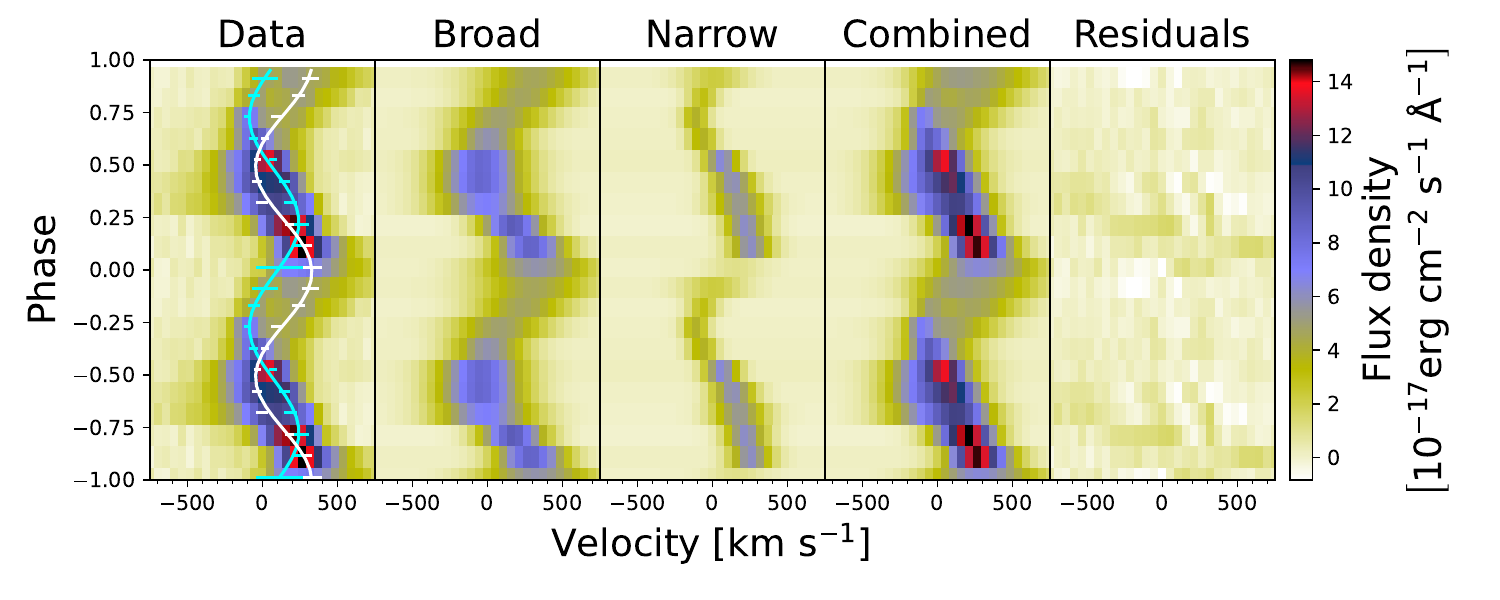}
\caption{Trailed spectra of H$\alpha$ (top) and H$\beta$ (bottom) 
emission line and the results of their fitting. Two orbital cycles are 
presented for clarity. From left to right: data, the broad component fit, 
the narrow component fit, the combination of the two components, and the residuals (given as data minus model). White and cyan error bars  show 1$\sigma$ 
uncertainties of the RVs of the broad and narrow components, while solid 
lines represent their best-fitting sine functions. }
\label{fig:Halpha_fit}
\end{figure*}

In Fig.~\ref{fig:trailed_full}, we demonstrate the trailed spectra folded
with the orbital period for each grism. The emission lines clearly show 
a sine-like variability of their positions, depending on radial velocities (RVs) of the emitting matter, with the orbital phase. These variations look
similar for different lines.
The lines have 
highly variable asymmetric profiles
indicating the presence  of at least two emission components (Fig.~\ref{fig:lines_phase}). Such profiles with two or three 
components  are typical 
for polars \citep[e.g.][]{1997A&A...319..894S}.  
 Accounting for our limited spectral resolution,  we tried to fit profiles in each phase bin with a combination of two Gaussian functions.
To do this, we used the Markov chain Monte Carlo procedure 
\citep{Foreman_Mackey_2013}. We  
found that one Gaussian 
is narrower and less intense (`narrow')  than the other one (`broad'). 
Some examples of the fit results are presented in Fig.~\ref{fig:lines_phase}.  As seen, the components are most clearly resolved near the orbital phase $\sim$ 0.8. 
Variations of the best-fitting parameters with the orbital phase for 
the H$\alpha$ line are shown in Fig.~\ref{fig:lines_fit_pars}.
 We than fitted RVs of each component by the sine function:
\begin{equation}
\upsilon_i(t) = \gamma_i + K_i \sin \left(  \frac { 2\pi(t-t_0)}{P_{\rm orb}} - \Delta \phi_i  \right),\ i \in \left\{ \text{narrow,\ broad} \right\},
\label{eq:sine}
\end{equation}
where  $t$ is time, $\gamma_i$ is the systemic RV, $K_i$ is the RV semi-amplitude, $\Delta \phi_i$ is the phase shift relative to the time $t_0$. 
The results for H$\alpha$ and H$\beta$ lines are presented in 
Fig.~\ref{fig:lines_fit_pars}, Fig.~\ref{fig:Halpha_fit} and 
Table~\ref{tab:sin-fit}. 

Both components of Balmer lines demonstrate  RVs curves with practically similar amplitudes but shifted from one to another in orbital phase by about 0.25.
Their peak intensities and FWHMs only marginally vary  with the orbital 
phase at  $\sim 1\sigma$ level. A high degeneracy between these parameters 
thus precludes any reliable analysis of their orbital variations.  
At the same time, maxima  of full  Balmer line intensities are 
reached at the phase of the maximum of the object photometric brightness ($\phi \approx 0.1$; Fig.~\ref{fig:Halpha_fit}).

\ion{He}{} lines are much weaker than the Balmer ones, and therefore, their fit results are too uncertain.

In polars, narrow Balmer, \ion{He}{I}, and \ion{He}{II} lines  likely arise
from the tip of the Roche lobe of the donor star and possibly  the beginning 
of the accretion stream, while the broad ones trace the accretion regions 
near  WD magnetic poles \citep{Schwope2011}. In addition,  the donor  is 
irradiated by the WD. This is typically observed via \ion{Na}{I} 8183,
8195 dublet and \ion{Ca}{II} 8498 and 8542 emission lines 
above the accretion continuum. They are formed closer to the donor centre 
of mass, however, they are located outside our observed range. Artistic 
representations of the system at three specific orbital phases are shown at the top of Fig.~\ref{fig:lines_fit_pars}. At phase 0.0 the WD magnetic polar region hot spot is located on the line of sight of the observer. 

%
%
\begin{table}
\caption{Best-fitting parameters with 1$\sigma$ uncertainties for the RV curves of spectral line components.}
\centering
\label{tab:sin-fit}
\begin{tabular}{ccccc}
\hline
Line      &  Component & $\gamma$, km~s$^{-1}$ & $K$, km~s$^{-1}$   & $\Delta \phi$ \\
\hline
H$\alpha$ & broad      & $141\pm25$            &  $221\pm36$  & $0.25\pm0.02$ \\
          & narrow     & $100\pm32$            & $190\pm41$         & $0.01\pm0.02$ \\ 
\hline
H$\beta$  & broad      & $140^{+31}_{-30}$     & $184^{+43}_{-42}$  & $0.26\pm0.04$ \\
          & narrow     & $100\pm25$            & $173^{+112}_{-98}$ & $0.04\pm0.05$ \\ 
\hline
\end{tabular}
\label{tab:sine_pars}
\end{table}

\subsection{System parameters}
 \label{SysPar}
 
 The light curves and spectral properties allow us to state that \src\ is likely a polar. To estimate the system parameters, 
we thus accepted the WD mass $M_1=0.8$~\msun\  which is the average value for mCVs 
\citep{2020MNRAS.498.3457S}. From the mass-period and the radius-mass 
relationships given by~\citet{Smith:1998}
\begin{align}
    \frac{M_{\rm 2}}{M_{\odot}} &= (0.126 \pm 0.011)P_{\rm orb} - (0.110 \pm 0.040), \\
    \frac{R_{\rm 2}}{R_{\odot}} &= (0.117 \pm 0.004)P_{\rm orb} - (0.011 \pm 0.018),
    \label{eq:massRadio}
\end{align}
where \pb\ represents the orbital period in hours, we estimated the 
secondary star's mass   $M_{\rm 2}=\mathrm{0.09\pm0.05~M_{\odot}}$ and 
its radius  $\mathrm {R_{\rm 2}=0.17\pm0.02~{\rm R}_{\odot}}$.  
This gives the mass ratio $M_2/M_1 \approx 0.1$.~
\src\ shows high optical variability, but no WD eclipses are seen in the 
light curves. Assuming that the H$\alpha$ narrow component 
is formed near the tip of the Roche lobe (L$_1$ point), given its RV  
semi-amplitude (Table~\ref{tab:sin-fit}) and the masses of the components 
we estimate  the system inclination to be $i = 48^\circ \pm 12^\circ$.
As we have mentioned above, the heating of the secondary star by high-energy
radiation emitted from the accreting magnetic white dwarf results in 
formation of emission lines closer to the donor centre of mass 
\citep{1992MNRAS.257..476D}.  Assuming that the narrow component velocity 
corresponds to the centre-of-mass velocity, we obtain the lower limit to the 
system inclination of about 36$^\circ$. 
  
Based on the system inclination and  duration of the plateau
state at maximum brightness $\Delta \phi \approx 0.3 P_{\rm orb}$, we estimated the co-latitude of the accretion spot at WD surface as
\begin{equation}
\beta = \arctan \left(-\frac{\text{cos}(\pi \Delta \phi)}{\text{tan}(i)} \right) = 28^\circ\pm10^\circ.
\end{equation}
It is in agreement with what is observed in single-pole accreting polars, where the co-latitude $\beta$ of the accretion spot typically lies in the range 20--50 degrees \citep{1988MNRAS.231..597C}.

\begin{figure*}
\centering
\includegraphics[width=1\linewidth]{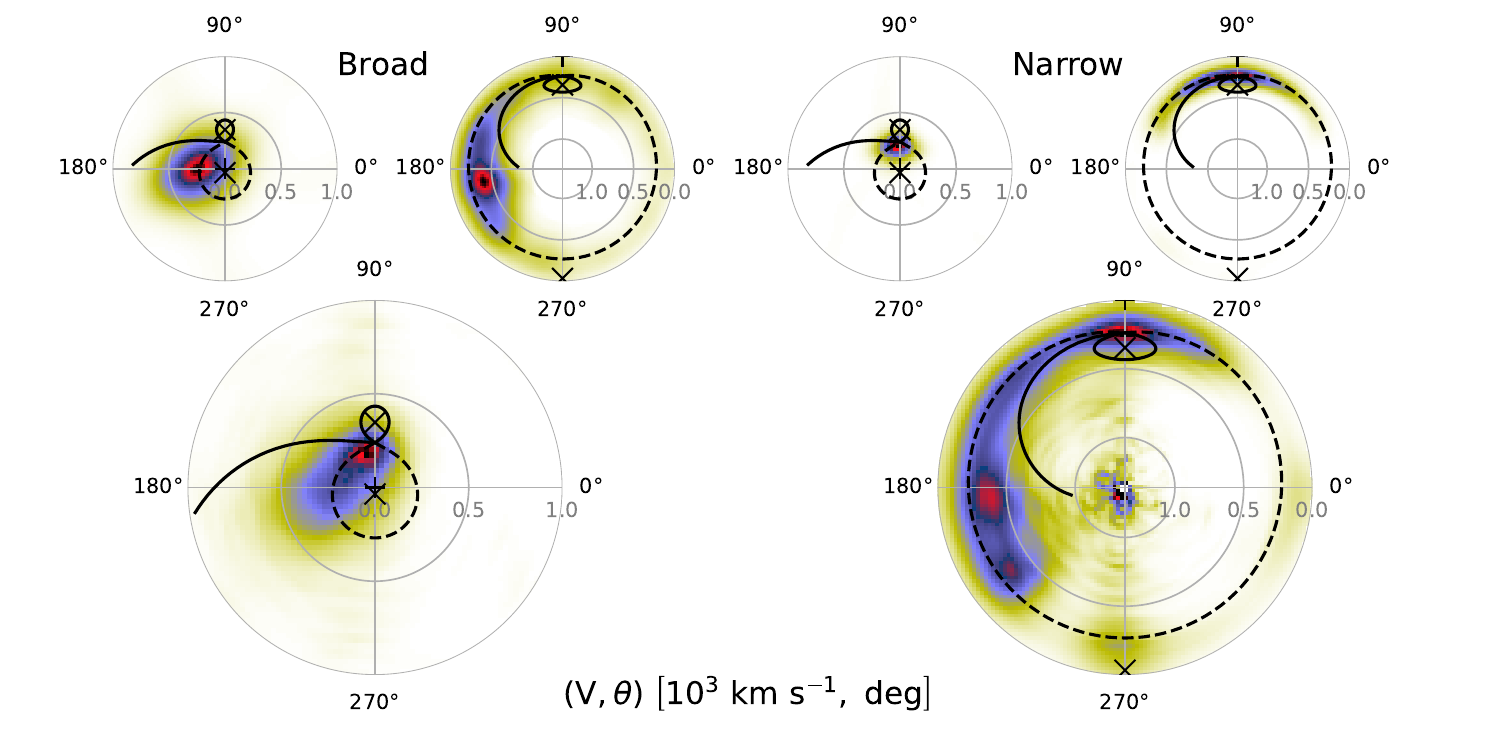} 
\caption{Doppler tomography of the H$\alpha$ line (bottom) and its 
components (top). The left panels demonstrate the standard Doppler 
tomograms while the right panels -- the inside-out projections.
The Roche lobes of the WD and companion (black dashed and solid closed 
lines, respectively) and the ballistic trajectory (the black curved line) 
are overlaid. The centre of mass of the system is marked by the cross 
while that of the WD and companion -- by the `$\times$' symbols. The model of the 
binary system is calculated assuming the white dwarf mass $M_1=0.8$~\msun, 
the mass ratio $M_2/M_1=0.1$ and the inclination $i=48$\degs.}
\label{fig:tomograms_Ha}
\end{figure*}
\begin{figure*}
\centering
\includegraphics[width=0.63\linewidth]{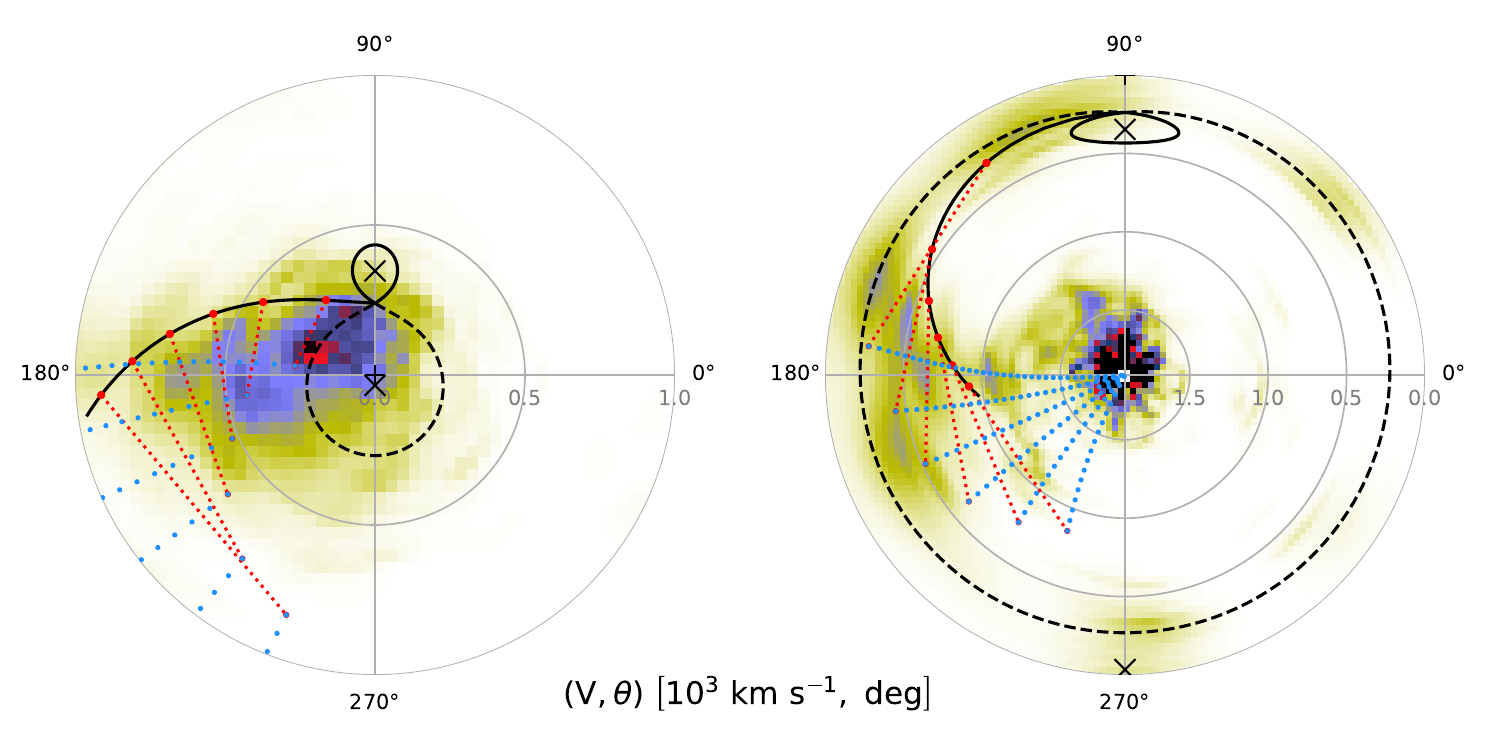} 
\includegraphics[width=0.33\linewidth, bb = 0 70 650 690, clip=]{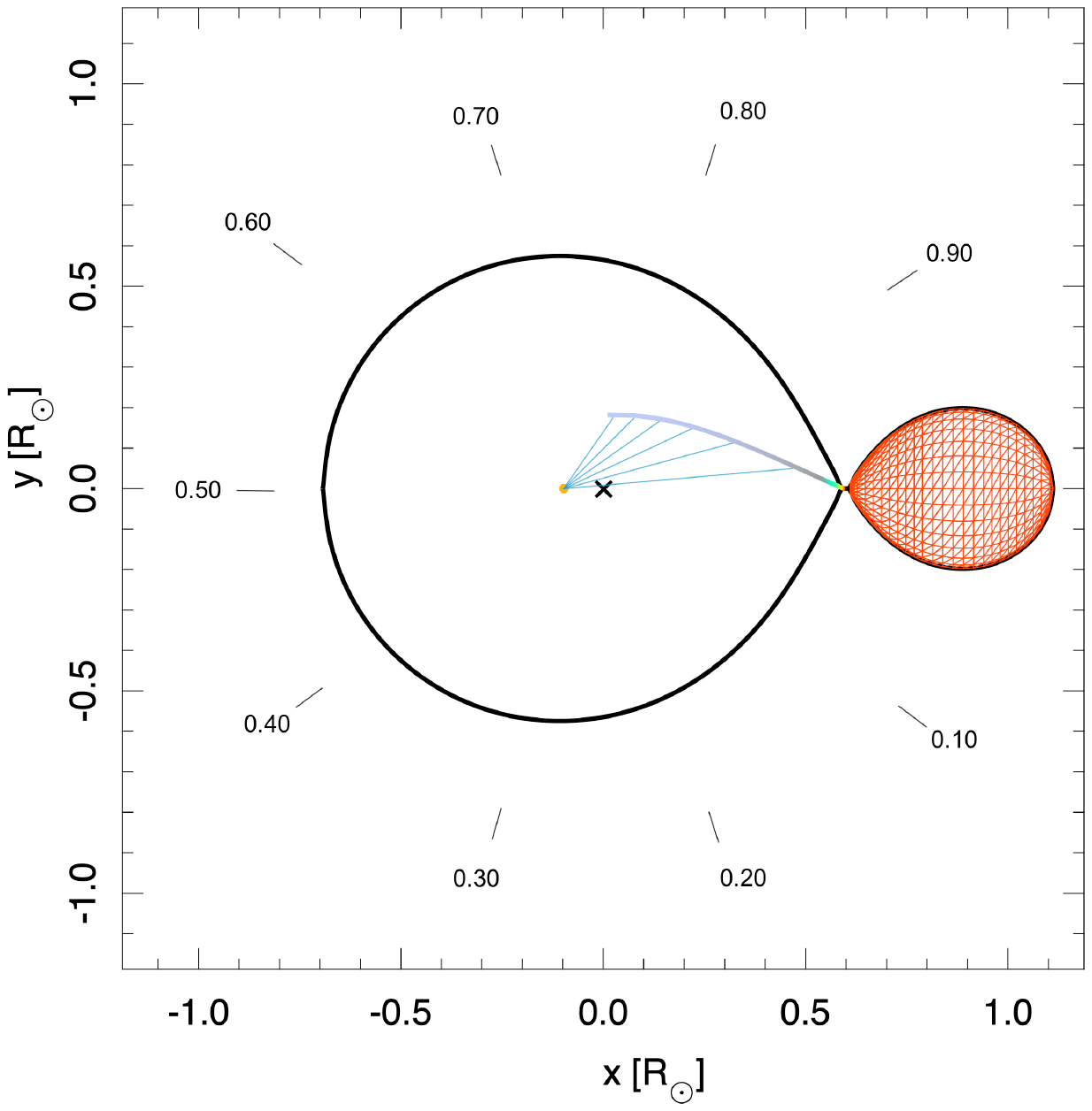}
\caption{Left and middle panels: Doppler tomography of the \ion{He}{ii}~4686 line. 
The left panel demonstrates the standard Doppler tomogram, while the middle
panel -- the inside-out projection. The designations are the same as in 
Fig.~\ref{fig:tomograms_Ha}. 
The dotted lines are magnetic trajectories (see text). 
Right panel: configuration of the system in spatial coordinates. 
The magnetic dipole field lines at 10$^\circ$ intervals from 5$^\circ$ to 55$^\circ$ in azimuth around the primary are shown for an example. We assumed the dipole axis azimuth and co-latitude are 27$^\circ$ and 45$^\circ$.
The symbol `$\times$' shows the binary system's centre of mass. The orbital phases are marked within the plot.}

\label{fig:tomograms_He}
\end{figure*}

\begin{figure}
\begin{minipage}[h]{1.\linewidth}
\center{
\includegraphics[width=0.925\linewidth,clip=]{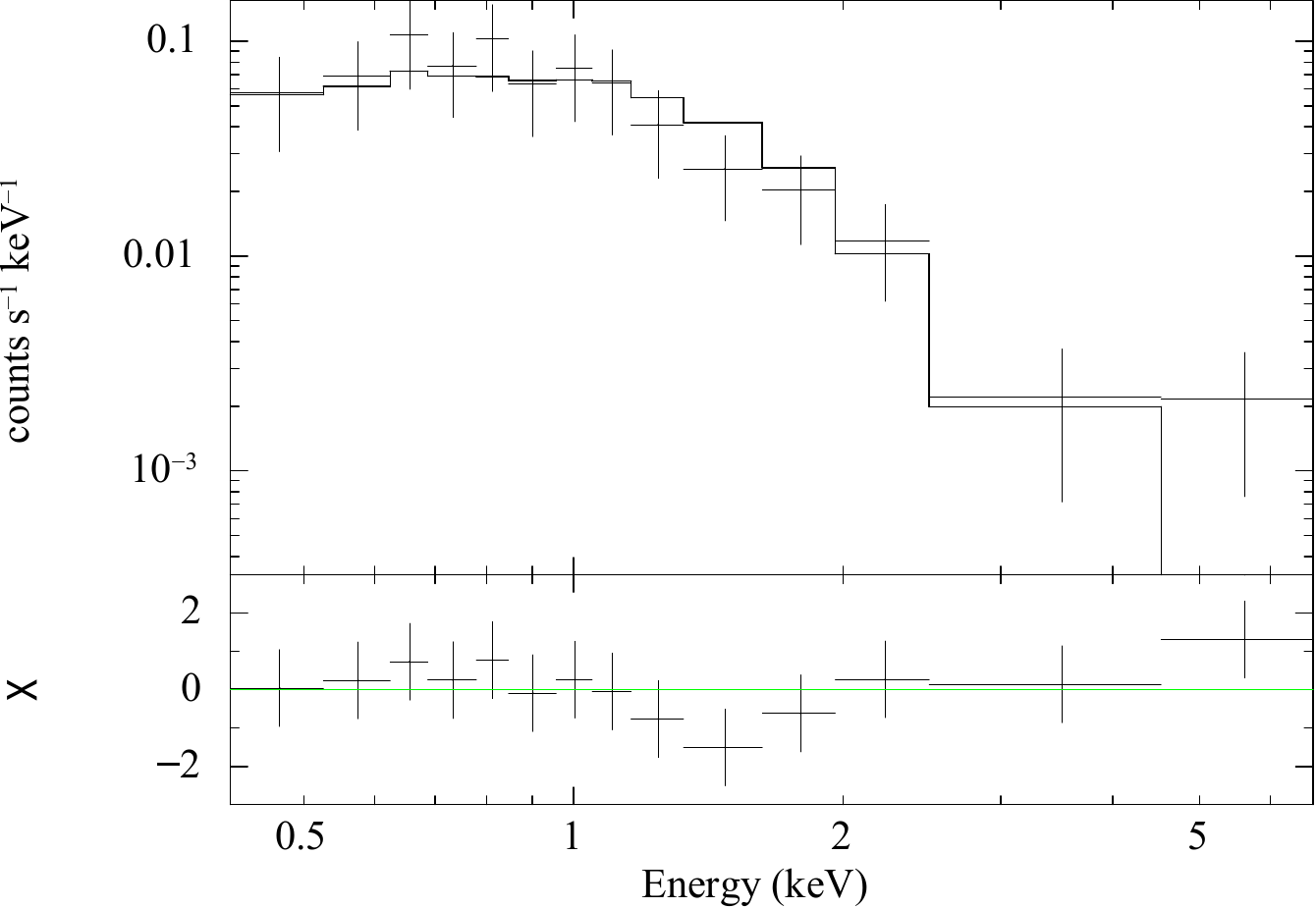}
}
\end{minipage}
\caption{\textit{Top}: the X-ray spectrum of \src\ (crosses) and the 
best-fitting model {\sc tbabs}$\times${\sc mekal} (solid line). 
For illustrative purposes, the spectrum was grouped to ensure at 
least 6 counts per bin. \textit{Bottom}: residuals.}
\label{fig:xray-spec}
\end{figure}

\subsection{Doppler tomography}
\label{subsec:tomography}

We used the Doppler tomography technique \citep{Marsh1988} to probe the 
accretion flow structure in \src. This technique is a well-known method 
to map the emission of gas moving in the velocity space ($V$,~$\theta$), where $V$ is the absolute value of the velocity measured relative to the 
system's centre of mass and $\theta$ is the angle between the velocity 
vector and the line between the centers of mass of the primary and 
secondary stars. We generated Doppler maps from  the GTC time-resolved 
spectra using the maximum entropy method \citep{spruit1998} 
as implemented by \citet{Kotze:2016}. 
The Roche lobe of the secondary and the primary, the positions of centres of mass of stars and the system,  and the ballistic trajectory of the accreted matter 
are calculated using
the system parameters. 
The results for the
H$\alpha$ line are presented in Fig.~\ref{fig:tomograms_Ha}. 
In the bottom panels of the figure, the standard and inside-out projection 
tomograms are displayed. 
They reveal a relatively bright compact and a more extended emission regions.
The zero phase is a free parameter  
and it was fixed 
to locate the emission maximum of the 
compact region 
at the L$_1$ point, i.e. at the beginning 
of the ballistic stream, as observed in eclipsing polars \citep{1997A&A...319..894S, 2002MNRAS.331..488S}. 
This gives  $t_0$~=~HJD 2460617.620472.   
In this case, the more extended emission is distributed  
between the ballistic trajectory and the boundary of the WD Roche lobe, 
and the maximum of its intensity is shifted by $\approx$0.25$P_{\rm orb}$ in
respect to the L$_1$ point. A similar picture is obtained for other Balmer as well as \ion{He}{i} lines.

To understand the origin of the two emission regions, we constructed model tomograms for H$\alpha$ narrow and broad components separately 
based on the result of the RV fit. They are presented in the top panels of 
Fig.~\ref{fig:tomograms_Ha}.
As expected,  the narrow line component corresponds to the compact bright region 
at the L$_1$ point ($V\approx 200$ km~s$^{-1}$, $\theta  \approx 90^\circ$). 
The broad component is extended along the Roche lobe boundary away from the stream ballistic trajectory and has a compact brightest excess at $V\approx 250$ km~s$^{-1}$, $\theta  \approx 190^\circ$.

In contrast to Balmer and \ion{He}{i} lines, no emission spot is seen 
near the L$_1$ point for the \ion{He}{ii}~4686 line tomogram 
(left and middle panels of Fig.~\ref{fig:tomograms_He}).
For illustrative purposes, we overlay an example of magnetic trajectories following \citet{Kotze:2016}.
We also show the system in spatial coordinates (Fig.~\ref{fig:tomograms_He}, right). 

\subsection{X-ray spectrum}
\label{subsec:xray-spec}

Utilising the eSASS tasks, we extracted the \src\ spectra from the \eros\ 
data in a circular region with a 60-arcsec radius. For the background
extraction, we chose an annulus region around the source with the inner 
and outer radii of 150 and 300 arcsec. We obtained 79 net counts and 
grouped the spectra to ensure at least 1 count per bin.

The spectra were fitted in the 0.3--9 keV band with the X-Ray Spectral 
Fitting Package ({\sc xspec}) v.12.13.1 \citep{xspec}. Due to the low 
number of counts, we used the $W$-statistics appropriate for Poisson 
data with Poisson background \citep{wstat}. To account for the 
interstellar absorption, we applied the {\sc tbabs} model with the 
{\sc wilm} abundances \citep*{wilms2000}. According to the 3D dust map 
of \citet{dustmap2019}, the maximal reddening in the \src\ direction 
is low, $E(B-V)=0.05$ mag. We converted this value to the column 
density \nh\ using the empirical relation from \citet{foight2016}.
The resulting value, \nh~=~$4.4\times 10^{20}$ cm$^{-1}$, was fixed 
during the fitting procedure.

We found that the spectrum can be well described by the optically 
thin thermal plasma model {\sc mekal} \citep{mewe1985,mewe1986,liedal1995} 
with the temperature $T=7^{+25}_{-4}$ keV, the unabsorbed flux in the 0.3--10 keV band $F_X^{\rm 0.3-10\ keV}=5.1^{+2.1}_{-1.4}\times 10^{-13}$ \flux\ and $W=63$ for 82 degrees of freedom. 
The spectrum and the best-fitting model are presented in Fig.~\ref{fig:xray-spec}.

 
\section{Discussion and conclusions}
\label{sec:discussion}

\begin{figure}
\centering
\includegraphics[width=0.95\linewidth,bb = 0 16 600 625, clip=]{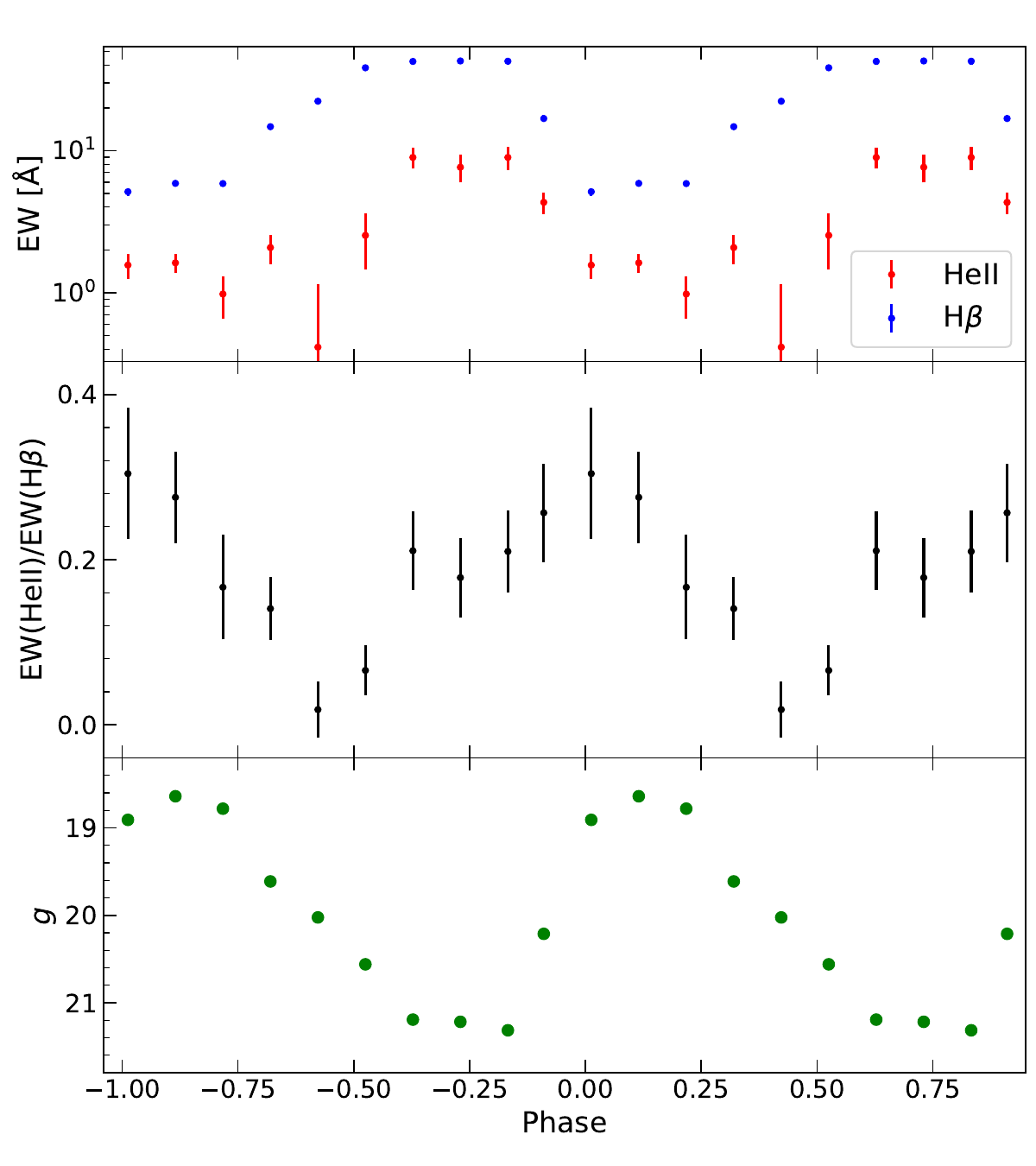}
\caption{Equivalent widths of the \ion{He}{ii} 4686 (red dots) 
and H$\beta$ (blue dots; top) emission lines, their ratio (middle) 
and intensity of spectra in the $g$ band (bottom) vs orbital phase.} 
\label{fig:EW-phase}
\end{figure}

\begin{figure}
    \centering
    \includegraphics[width=0.95\linewidth, bb = 0 10 450 300, clip=]{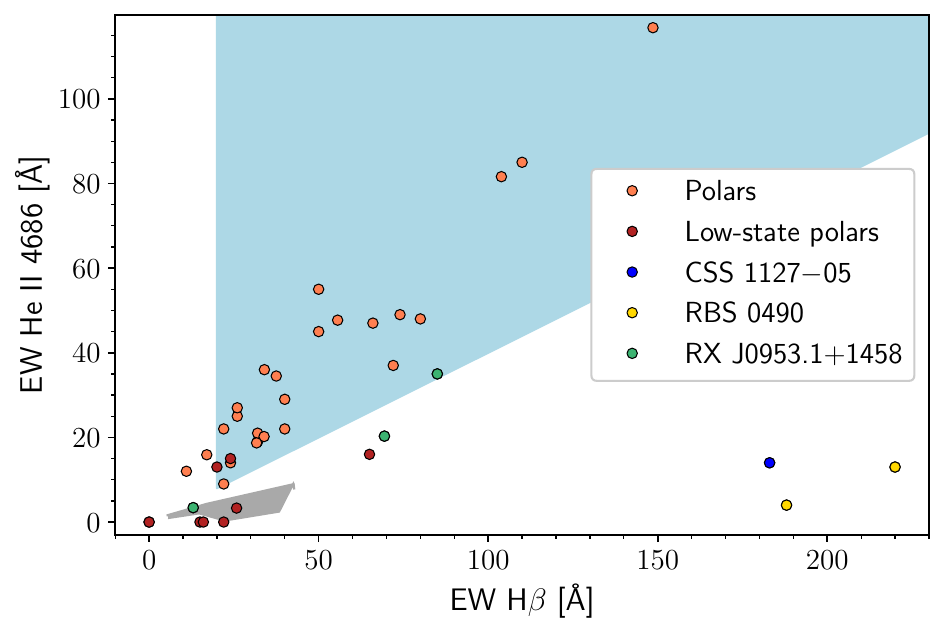}
    \caption{EWs of the \ion{He}{ii} 4686 vs H$\beta$ emission lines
    for different polars in high and low states (orange and dark-red 
    points). The blue area corresponds to the Silber's criterion for 
    mCVs. The grey region near the low-left corner shows the \src\ 
    position at different orbital phases (see Fig.~\ref{fig:EW-phase}). 
    Measurements for other sources are taken from \citet{oliveira2017, 
    oliveira2020, Beuermann2021, shafter1995, joshi2020, joshi2022, 
    Voikhanskaya1986, Singh1995, Griffiths1979, Romero-Colmenero2003, 
    Tovmassian1997}. Sources for which EW(\ion{He}{ii} 
    4686)/EW(H$\beta$) < 0.4 are marked in blue, yellow and green 
    (see text; several points of the same colour correspond to 
    different observations of a specific source). }
    \label{fig:He-vs-Hb}
\end{figure}

We have analysed optical and X-ray observations of \src.
Using optical photometry and time-resolved spectroscopy, we confirmed 
the orbital period and high amplitude of brightness modulation of 
$\approx$3 mag found by \citet{denisenko2018}. Archival optical data reveal the high and low brightness states of \src\, and it has been in the high state since at least 2018 until our observations. No evidence of eclipses or outbursts was found in the data covering the period from the end of 2023 to the beginning of 2025. 

The GTC optical spectra obtained in the high state demonstrate Balmer 
and \ion{He}{I} emission lines 
as well as unusually weak 
\ion{He}{II}.
The Doppler tomography revealed that \src\ is a system with an accretion stream instead of a disk which unambiguously confirms its classification as a polar and rejects the `spider' pulsar interpretation.
The emission lines have asymmetric shapes and show significant profile 
and RV variations with the orbital phase. The line profiles can be 
described by a model consisting of narrow and broad components associated with different emission regions. This is a common property 
of polars \citep[e.g.][]{Beuermann2021,Liu2023,Lin2025}. The RV curves 
of the components have similar amplitudes but are shifted by about 0.25 in 
phase. Such behaviour of Balmer emission lines was observed in the polar 
CP Tuc ($G=19.1$ mag, $D=207$ pc)  with an orbital period of 0.061832 d, close to that of \src\ 
\citep{1996A&A...315L...1T}. The WD or donor star are not resolved in 
our spectra. Cyclotron humps typical for polars are not detected either, 
and the continuum is smooth. It varies with the orbital period, having the maximum amplitude of variation at $\sim 6000$\AA. 
This is also similar to the case of CP Tuc, though the maximal variation of
its continuum is smaller ($\approx$2~mag) and occurs at $\approx 8800$~\AA.

Typically, spectra of polars show a strong \ion{He}{II}~4686 emission 
line with an equivalent width (EW) comparable to that of H$\beta$ 
\citep[e.g.][]{szkody1998}.
In Fig.~\ref{fig:EW-phase}, we present variations of EWs of \ion{He}
{II}~4686 and H$\beta$ lines\footnote{EWs are presented as positives.} 
as well as their ratio with the \src\ orbital phase.
One can see that the EWs vary in antiphase with the source brightness 
while the line ratio changes roughly in phase with it. 
This may indicate that at the maximum brightness phase the emission 
comes from the hotter plasma with a higher ionisation rate from the 
irradiated secondary's surface and/or the region where the stream 
interacts with the WD magnetosphere, forming a hot spot. 

\citet{silberphd} provided the following criterion for separating mCVs from non-magnetic CVs: EW(\ion{He}{II}~4686)/EW(H$\beta$)~$>$~0.4 and 
EW(H$\beta$)~$>$~20~\AA. 
For \src, the ratio EW(\ion{He}{II}~4686)/EW(H$\beta$) is $\lesssim$0.4 over the whole orbital period (Fig.~\ref{fig:EW-phase}, middle) with the averaged value of $\approx$0.2, which is unexpectedly low for mCVs.
A similar picture 
is seen in spectra of some low-accretion-state polars 
\citep[e.g.][]{lathan1981,kolbin2024}.  
However, \src\ 
was in the high state during our observations.  observations. 
There are a few high-state objects with  low EW(\ion{He}{II}~4686)/EW(H$\beta$) ratios. 
For example, \citet{Beuermann2021} and \citet{oliveira2020} obtained EW(\ion{He}{II}~4686)/EW(H$\beta$)~$\sim$~0.3--0.4 for the polar RX J0953.1+1458 with a rather low magnetic field (19 MG).
RBS 0490 proposed as a low-field polar shows very strong Balmer lines and has extremely low EW(\ion{He} {II}~4686)/EW(H$\beta$) = 0.02--0.06 \citep{joshi2022}. 
Similar small ratios were derived for the polar candidate CSS1127$-$05 \citep{oliveira2017} and the polar LSQ1725$-$64 with a low magnetic field (12MG) \citep{fuchs2016}.
In Fig.~\ref{fig:He-vs-Hb}, the data of EW(\ion{He}{II}~4686) vs. EW(H$\beta$) for mCVs are presented.
Unlike other high-state polars, \src\ does not satisfy Silber’s criterion (EW(\ion{He}{II}~4686)/EW(H$\beta$)~$>$~0.4) at any orbital phase. The reason for this unusually low ratio remains uncertain. 
The \ion{He}{II}~4686 emission  can be suppressed by a high magnetic field
when cyclotron cooling dominates over bremsstrahlung, and/or via a low accretion rate, which can also lead to a complete absence of \ion{He}{II} emission (see, e.g., \citealt{2010MNRAS.403..755K}).

The X-ray data for \src\ were obtained when it was in the high state.
The unabsorbed flux in the 0.2--2.3 keV band obtained from the \eros\ data 
is $F_X^{\rm 0.2-2.3\ keV} \approx 2.3\times10^{-13}$~\flux.
The flux presented in the XMMSLEWCLN catalogue is $\sim$6 times higher.
This is not surprising due to the strong variability of the source. The 
X-ray spectrum can be well fitted by the thermal plasma model with a 
temperature compatible with those obtained for other CVs 
\citep[e.g.][]{galiullin2024}. Taking these two
flux estimates and the optical variability $V\approx17.3$--20.5 mag, we 
calculated the X-ray to optical flux ratio log$(F_X/F_{\rm opt}) 
\sim$~$-0.3$--1.8. Although the X-ray observations do not cover the full 
orbital period and thus we do not know the exact X-ray flux variation, the 
derived ratio is in agreement with polars \citep{schwope2024, Lin2025}: 
the majority of polars have $\text{log} (F_X/F_{\rm opt})>-1.6$, and a 
substantial fraction -- $\text{log} (F_X/F_{\rm opt})>-0.3$.

The accretion rate $\dot{M}$ can be estimated using the bolometric X-ray 
luminosity $L_{X,\rm bol}$ via the standard formula
\begin{equation}
L_{X,\rm bol} \sim G M_1 \dot{M}/R_1, 
\end{equation}
where $M_1$ and $R_1$ are the WD mass and radius, and $G$ is the 
gravitational constant. This transforms into
\begin{equation}
\dot{M} \sim 1.4 \times 10^{-11} D^2 F_{X,\rm bol} R_1 / M_1 \ \ {\rm M_\odot~yr^{-1}}, 
\end{equation}
where the distance $D$ is expressed in kpc, the mass $M_1$ -- in Solar 
masses, the radius $R_1$ -- in 10000 km and the bolometric X-ray flux 
$F_{X,\rm bol}$ -- in 10$^{-12}$~\flux. From the \eros\ spectrum, we 
estimated $F_{X,\rm bol}  \sim 0.7$. 
The distance can be roughly constrained as follows.
As seen from Fig.~\ref{fig:panztf}, in the low state of  \src\  its minimal optical brightness is $r \sim 21.5$ mag. 
By analogy with other polars in the low states, we can assume that the companion dominates in the red part of the optical spectrum.
Based on the secondary mass and radius estimated in Sec.~\ref{SysPar}, 
we assigned it the M5V spectral type as the brightest among 
other possible later types of stars with similar masses and 
radii\footnote{\url{https://www.pas.rochester.edu/~emamajek/EEM_dwarf\_UBVIJHK\_colors\_Teff.txt}}. 
The absolute magnitude of the M5V type is $M_V = 14.15$.
Using the reddening \src\ $E(B-V)=0.05$ mag and the transformation equation 
$r = V - 0.49(B-V) + 0.11$ \citep{sdss_bessel_convertion} with the colour 
index $B-V = 1.83$, corresponding to the alleged spectral type, we estimated 
the absolute magnitude $M_r = 13.3$. This results in a distance of 
$\sim$400 pc and $\dot{M} \sim 2\times10^{-12}$~\msun~yr$^{-1}$.
The latter is somewhat lower than that observed for polars in high states 
\citep[e.g.][]{Beuermann2021}. However, this value should be considered
with caution, given the large X-ray flux and distance uncertainties.  
If the contribution from the secondary to the optical brightness is lower 
than 21.5~mag, then the distance and $\dot{M}$ would be larger. 
However, the spectral type can be later, e.g. M6V--M9V, with a
larger $M_V$ providing a lower distance.

Although we confirmed that \src\ is a polar, it is not possible to
accurately measure the parameters of the binary components and the distance 
using the current data. The distance will be potentially provided in the 
future \gaia\ release. Monitoring of \src\ in the optical to reveal it in
the low state would be useful to estimate the masses and temperatures of 
the components. High-resolution optical and infrared spectral observations 
are needed to search for Zeeman splitting in absorption lines and cyclotron 
humps in the continuum emission and thus to measure the WD magnetic field.
Deeper X-ray observations would provide the orbital flux modulation in this 
range and help to better constrain the accretion rate. Polarimetric 
observations are necessary to clarify the system geometry and the magnetic field structure. 


\section*{Acknowledgements}

We thank the anonymous referee for useful comments.
The work is based on observations made with the GTC telescope, in the Spanish Observatorio del Roque de los Muchachos of the Instituto de Astrof\'isica de Canarias, under Director’s Discretionary Time and the Observatorio Astron\'omico Nacional on the Sierra San Pedro M\'artir (OAN-SPM), Baja California, M\'exico. We thank the daytime and night support staff at the OAN-SPM for facilitating and helping obtain our observations.
Based on observations obtained with the Samuel Oschin Telescope 48-inch and the 60-inch Telescope at the Palomar Observatory as part of the Zwicky Transient Facility project. ZTF is supported by the National Science Foundation under Grants No. AST-1440341 and AST-2034437 and a collaboration including current partners Caltech, IPAC, the Oskar Klein Center at Stockholm University, the University of Maryland, University of California, Berkeley, the University of Wisconsin at Milwaukee, University of Warwick, Ruhr University, Cornell University, Northwestern University and Drexel University. Operations are conducted by COO, IPAC, and UW.
The CSS survey is funded by the National Aeronautics and Space Administration under Grant No. NNG05GF22G issued through the Science Mission Directorate Near-Earth Objects Observations Program. The CRTS survey is supported by the U.S.~National Science Foundation under grants AST-0909182 and AST-1313422.
This work used data obtained with eROSITA telescope onboard SRG observatory. The SRG observatory was built by Roskosmos in the interests of the Russian Academy of Sciences represented by its Space Research Institute (IKI) in the framework of the Russian Federal Space Program, with the participation of the Deutsches Zentrum f\"{u}r Luft- und Raumfahrt (DLR). The SRG/eROSITA X-ray telescope was built by a consortium of German Institutes led by MPE, and supported by DLR. The SRG spacecraft was designed, built, launched and is operated by the Lavochkin Association and its subcontractors. The science data are downlinked via the Deep Space Network Antennae in Bear Lakes, Ussurijsk, and Baykonur, funded by Roskosmos. The eROSITA data used in this work were processed using the eSASS software system developed by the German eROSITA consortium and proprietary data reduction and analysis software developed by the Russian eROSITA Consortium.
The work of AVB, AVK, DAZ and YAS was supported by the baseline project FFUG-2024-0002 of the Ioffe Institute (analysis of archival optical data, X-ray data and GTC data obtained on 2024-11-03). 
SVZ acknowledges the DGAPA-PAPIIT grant IN119323. 
AYK acknowledges the DGAPA-PAPIIT grant IA105024. 
The work is partially carried out within the framework of the Project No. BR20280974 "Program of fundamental astrophysical research in Kazakhstan: observations and theory", financed by the Ministry of Education and Science of the Republic of Kazakhstan.

\section*{Data Availability}

The spectroscopic data are available through the GTC data archive: 
\url{https://gtc.sdc.cab.inta-csic.es/gtc/}, CSS data -- 
\url{http://nesssi.cacr.caltech.edu/DataRelease/}, \ps\ data -- 
\url{https://catalogs.mast.stsci.edu/panstarrs/}, ZTF data -- 
\url{https://irsa.ipac.caltech.edu/Missions/ztf.html}, OAN-SPM,
MAO, ATO and \eros\ data -- upon request.



\bibliographystyle{mnras}
\bibliography{ref} 

\begin{thebibliography}{}
\makeatletter
\relax
\def\mn@urlcharsother{\let\do\@makeother \do\$\do\&\do\#\do\^\do\_\do\%\do\~}
\def\mn@doi{\begingroup\mn@urlcharsother \@ifnextchar [ {\mn@doi@}
  {\mn@doi@[]}}
\def\mn@doi@[#1]#2{\def\@tempa{#1}\ifx\@tempa\@empty \href
  {http://dx.doi.org/#2} {doi:#2}\else \href {http://dx.doi.org/#2} {#1}\fi
  \endgroup}
\def\mn@eprint#1#2{\mn@eprint@#1:#2::\@nil}
\def\mn@eprint@arXiv#1{\href {http://arxiv.org/abs/#1} {{\tt arXiv:#1}}}
\def\mn@eprint@dblp#1{\href {http://dblp.uni-trier.de/rec/bibtex/#1.xml}
  {dblp:#1}}
\def\mn@eprint@#1:#2:#3:#4\@nil{\def\@tempa {#1}\def\@tempb {#2}\def\@tempc
  {#3}\ifx \@tempc \@empty \let \@tempc \@tempb \let \@tempb \@tempa \fi \ifx
  \@tempb \@empty \def\@tempb {arXiv}\fi \@ifundefined
  {mn@eprint@\@tempb}{\@tempb:\@tempc}{\expandafter \expandafter \csname
  mn@eprint@\@tempb\endcsname \expandafter{\@tempc}}}

\bibitem[\protect\citeauthoryear{{Arnaud}}{{Arnaud}}{1996}]{xspec}
{Arnaud} K.~A.,  1996, in {Jacoby} G.~H.,  {Barnes} J.,  eds,  Astronomical
  Society of the Pacific Conference Series Vol. 101, Astronomical Data Analysis
  Software and Systems V. p.~17

\bibitem[\protect\citeauthoryear{{Beuermann}, {Burwitz}, {Reinsch}, {Schwope}
  \& {Thomas}}{{Beuermann} et~al.}{2021}]{Beuermann2021}
{Beuermann} K.,  {Burwitz} V.,  {Reinsch} K.,  {Schwope} A.,   {Thomas} H.~C.,
  2021, \mn@doi [\aap] {10.1051/0004-6361/202038598}, \href
  {https://ui.adsabs.harvard.edu/abs/2021A&A...645A..56B} {645, A56}

\bibitem[\protect\citeauthoryear{{Bobakov}, {Kirichenko}, {Zharikov},
  {Karpova}, {Zyuzin}, {Shibanov}, {Mennickent}  \&
  {Garcia-{\'A}lvarez}}{{Bobakov} et~al.}{2024}]{Bobakov2024}
{Bobakov} A.~V.,  {Kirichenko} A.~Y.,  {Zharikov} S.~V.,  {Karpova} A.~V.,
  {Zyuzin} D.~A.,  {Shibanov} Y.~A.,  {Mennickent} R.~E.,
  {Garcia-{\'A}lvarez} D.,  2024, \mn@doi [\aap] {10.1051/0004-6361/202450205},
  \href {https://ui.adsabs.harvard.edu/abs/2024A&A...690A.173B} {690, A173}

\bibitem[\protect\citeauthoryear{{Brunner} et~al.,}{{Brunner}
  et~al.}{2022}]{Brunner2022}
{Brunner} H.,  et~al., 2022, \mn@doi [\aap] {10.1051/0004-6361/202141266},
  \href {https://ui.adsabs.harvard.edu/abs/2022A&A...661A...1B} {661, A1}

\bibitem[\protect\citeauthoryear{{Cropper}}{{Cropper}}{1988}]{1988MNRAS.231..597C}
{Cropper} M.,  1988, \mn@doi [\mnras] {10.1093/mnras/231.3.597}, \href
  {https://ui.adsabs.harvard.edu/abs/1988MNRAS.231..597C} {231, 597}

\bibitem[\protect\citeauthoryear{{Cropper}}{{Cropper}}{1990}]{cropper1990}
{Cropper} M.,  1990, \mn@doi [\ssr] {10.1007/BF00177799}, \href
  {https://ui.adsabs.harvard.edu/abs/1990SSRv...54..195C} {54, 195}

\bibitem[\protect\citeauthoryear{{Davey} \& {Smith}}{{Davey} \&
  {Smith}}{1992}]{1992MNRAS.257..476D}
{Davey} S.,  {Smith} R.~C.,  1992, \mn@doi [\mnras] {10.1093/mnras/257.3.476},
  \href {https://ui.adsabs.harvard.edu/abs/1992MNRAS.257..476D} {257, 476}

\bibitem[\protect\citeauthoryear{{Denisenko}}{{Denisenko}}{2018}]{denisenko2018}
{Denisenko} D.,  2018, The Astronomer's Telegram, \href
  {https://ui.adsabs.harvard.edu/abs/2018ATel11626....1D} {11626, 1}

\bibitem[\protect\citeauthoryear{{Draghis}, {Romani}, {Filippenko}, {Brink},
  {Zheng}, {Halpern}  \& {Camilo}}{{Draghis} et~al.}{2019}]{draghis2019}
{Draghis} P.,  {Romani} R.~W.,  {Filippenko} A.~V.,  {Brink} T.~G.,  {Zheng}
  W.,  {Halpern} J.~P.,   {Camilo} F.,  2019, \mn@doi [\apj]
  {10.3847/1538-4357/ab378b}, \href
  {https://ui.adsabs.harvard.edu/abs/2019ApJ...883..108D} {883, 108}

\bibitem[\protect\citeauthoryear{{Drake} et~al.,}{{Drake} et~al.}{2009}]{drake}
{Drake} A.~J.,  et~al., 2009, \mn@doi [\apj] {10.1088/0004-637X/696/1/870},
  \href {https://ui.adsabs.harvard.edu/abs/2009ApJ...696..870D} {696, 870}

\bibitem[\protect\citeauthoryear{{Flewelling} et~al.,}{{Flewelling}
  et~al.}{2020}]{ps2020}
{Flewelling} H.~A.,  et~al., 2020, \mn@doi [\apjs] {10.3847/1538-4365/abb82d},
  \href {https://ui.adsabs.harvard.edu/abs/2020ApJS..251....7F} {251, 7}

\bibitem[\protect\citeauthoryear{{Foight}, {G{\"u}ver}, {{\"O}zel}  \&
  {Slane}}{{Foight} et~al.}{2016}]{foight2016}
{Foight} D.~R.,  {G{\"u}ver} T.,  {{\"O}zel} F.,   {Slane} P.~O.,  2016,
  \mn@doi [\apj] {10.3847/0004-637X/826/1/66}, \href
  {https://ui.adsabs.harvard.edu/abs/2016ApJ...826...66F} {826, 66}

\bibitem[\protect\citeauthoryear{Foreman-Mackey, Hogg, Lang  \&
  Goodman}{Foreman-Mackey et~al.}{2013}]{Foreman_Mackey_2013}
Foreman-Mackey D.,  Hogg D.~W.,  Lang D.,   Goodman J.,  2013, \mn@doi
  [Publications of the Astronomical Society of the Pacific] {10.1086/670067},
  125, 306–312

\bibitem[\protect\citeauthoryear{{Fuchs} et~al.,}{{Fuchs}
  et~al.}{2016}]{fuchs2016}
{Fuchs} J.~T.,  et~al., 2016, \mn@doi [\mnras] {10.1093/mnras/stw1759}, \href
  {https://ui.adsabs.harvard.edu/abs/2016MNRAS.462.2382F} {462, 2382}

\bibitem[\protect\citeauthoryear{{Fukugita}, {Ichikawa}, {Gunn}, {Doi},
  {Shimasaku}  \& {Schneider}}{{Fukugita}
  et~al.}{1996}]{sdss_bessel_convertion}
{Fukugita} M.,  {Ichikawa} T.,  {Gunn} J.~E.,  {Doi} M.,  {Shimasaku} K.,
  {Schneider} D.~P.,  1996, \mn@doi [\aj] {10.1086/117915}, \href
  {https://ui.adsabs.harvard.edu/abs/1996AJ....111.1748F} {111, 1748}

\bibitem[\protect\citeauthoryear{{Gaia Collaboration} et~al.,}{{Gaia
  Collaboration} et~al.}{2016}]{gaia2016}
{Gaia Collaboration} et~al., 2016, \mn@doi [\aap]
  {10.1051/0004-6361/201629272}, \href
  {https://ui.adsabs.harvard.edu/abs/2016A&A...595A...1G} {595, A1}

\bibitem[\protect\citeauthoryear{{Gaia Collaboration} et~al.,}{{Gaia
  Collaboration} et~al.}{2023}]{gaia-dr3-2023}
{Gaia Collaboration} et~al., 2023, \mn@doi [\aap]
  {10.1051/0004-6361/202243940}, \href
  {https://ui.adsabs.harvard.edu/abs/2023A&A...674A...1G} {674, A1}

\bibitem[\protect\citeauthoryear{{Galiullin} et~al.,}{{Galiullin}
  et~al.}{2024}]{galiullin2024}
{Galiullin} I.,  et~al., 2024, \mn@doi [\aap] {10.1051/0004-6361/202450734},
  \href {https://ui.adsabs.harvard.edu/abs/2024A&A...690A.374G} {690, A374}

\bibitem[\protect\citeauthoryear{{Green}, {Schlafly}, {Zucker}, {Speagle}  \&
  {Finkbeiner}}{{Green} et~al.}{2019}]{dustmap2019}
{Green} G.~M.,  {Schlafly} E.,  {Zucker} C.,  {Speagle} J.~S.,   {Finkbeiner}
  D.,  2019, \mn@doi [\apj] {10.3847/1538-4357/ab5362}, \href
  {https://ui.adsabs.harvard.edu/abs/2019ApJ...887...93G} {887, 93}

\bibitem[\protect\citeauthoryear{{Griffiths}, {Ward}, {Blades}, {Wilson},
  {Chaisson}  \& {Johnston}}{{Griffiths} et~al.}{1979}]{Griffiths1979}
{Griffiths} R.~E.,  {Ward} M.~J.,  {Blades} J.~C.,  {Wilson} A.~S.,  {Chaisson}
  L.,   {Johnston} M.~D.,  1979, \mn@doi [\apjl] {10.1086/183030}, \href
  {https://ui.adsabs.harvard.edu/abs/1979ApJ...232L..27G} {232, L27}

\bibitem[\protect\citeauthoryear{{Joshi}, {Pandey}, {Raj}, {Singh}, {Anupama}
  \& {Singh}}{{Joshi} et~al.}{2020}]{joshi2020}
{Joshi} A.,  {Pandey} J.~C.,  {Raj} A.,  {Singh} K.~P.,  {Anupama} G.~C.,
  {Singh} H.~P.,  2020, \mn@doi [\mnras] {10.1093/mnras/stz3016}, \href
  {https://ui.adsabs.harvard.edu/abs/2020MNRAS.491..201J} {491, 201}

\bibitem[\protect\citeauthoryear{{Joshi}, {Pandey}, {Rawat}, {Raj}, {Wang}  \&
  {Singh}}{{Joshi} et~al.}{2022}]{joshi2022}
{Joshi} A.,  {Pandey} J.~C.,  {Rawat} N.,  {Raj} A.,  {Wang} W.,   {Singh}
  H.~P.,  2022, \mn@doi [\aj] {10.3847/1538-3881/ac6026}, \href
  {https://ui.adsabs.harvard.edu/abs/2022AJ....163..221J} {163, 221}

\bibitem[\protect\citeauthoryear{{Kafka}, {Tappert}  \& {Honeycutt}}{{Kafka}
  et~al.}{2010}]{2010MNRAS.403..755K}
{Kafka} S.,  {Tappert} C.,   {Honeycutt} R.~K.,  2010, \mn@doi [\mnras]
  {10.1111/j.1365-2966.2009.16063.x}, \href
  {https://ui.adsabs.harvard.edu/abs/2010MNRAS.403..755K} {403, 755}

\bibitem[\protect\citeauthoryear{{Kolbin} et~al.,}{{Kolbin}
  et~al.}{2024}]{kolbin2024}
{Kolbin} A.~I.,  et~al., 2024, \mn@doi [Astronomy Letters]
  {10.1134/S1063773724700154}, \href
  {https://ui.adsabs.harvard.edu/abs/2024AstL...50..335K} {50, 335}

\bibitem[\protect\citeauthoryear{{Kotze}, {Potter}  \& {McBride}}{{Kotze}
  et~al.}{2016}]{Kotze:2016}
{Kotze} E.~J.,  {Potter} S.~B.,   {McBride} V.~A.,  2016, \mn@doi [\aap]
  {10.1051/0004-6361/201629120}, \href
  {https://ui.adsabs.harvard.edu/abs/2016A&A...595A..47K} {595, A47}

\bibitem[\protect\citeauthoryear{{Latham}, {Liebert}  \& {Steiner}}{{Latham}
  et~al.}{1981}]{lathan1981}
{Latham} D.~W.,  {Liebert} J.,   {Steiner} J.~E.,  1981, \mn@doi [\apj]
  {10.1086/158985}, \href
  {https://ui.adsabs.harvard.edu/abs/1981ApJ...246..919L} {246, 919}

\bibitem[\protect\citeauthoryear{{Liedahl}, {Osterheld}  \&
  {Goldstein}}{{Liedahl} et~al.}{1995}]{liedal1995}
{Liedahl} D.~A.,  {Osterheld} A.~L.,   {Goldstein} W.~H.,  1995, \mn@doi
  [\apjl] {10.1086/187729}, \href
  {https://ui.adsabs.harvard.edu/abs/1995ApJ...438L.115L} {438, L115}

\bibitem[\protect\citeauthoryear{{Lin} et~al.,}{{Lin} et~al.}{2025}]{Lin2025}
{Lin} J.,  et~al., 2025, \mn@doi [\aap] {10.1051/0004-6361/202452177}, \href
  {https://ui.adsabs.harvard.edu/abs/2025A&A...694A.112L} {694, A112}

\bibitem[\protect\citeauthoryear{{Liu}, {Hwang}, {Zakamska}  \&
  {Thorstensen}}{{Liu} et~al.}{2023}]{Liu2023}
{Liu} Y.,  {Hwang} H.-C.,  {Zakamska} N.~L.,   {Thorstensen} J.~R.,  2023,
  \mn@doi [\mnras] {10.1093/mnras/stad1156}, \href
  {https://ui.adsabs.harvard.edu/abs/2023MNRAS.522.2719L} {522, 2719}

\bibitem[\protect\citeauthoryear{{Lomb}}{{Lomb}}{1976}]{lomb1976}
{Lomb} N.~R.,  1976, \mn@doi [\apss] {10.1007/BF00648343}, \href
  {https://ui.adsabs.harvard.edu/abs/1976Ap&SS..39..447L} {39, 447}

\bibitem[\protect\citeauthoryear{{Marsh} \& {Horne}}{{Marsh} \&
  {Horne}}{1988}]{Marsh1988}
{Marsh} T.~R.,  {Horne} K.,  1988, \mn@doi [\mnras] {10.1093/mnras/235.1.269},
  \href {https://ui.adsabs.harvard.edu/abs/1988MNRAS.235..269M} {235, 269}

\bibitem[\protect\citeauthoryear{{Masci} et~al.,}{{Masci} et~al.}{2019}]{ztf}
{Masci} F.~J.,  et~al., 2019, \mn@doi [\pasp] {10.1088/1538-3873/aae8ac}, \href
  {https://ui.adsabs.harvard.edu/abs/2019PASP..131a8003M} {131, 018003}

\bibitem[\protect\citeauthoryear{{Mata S{\'a}nchez} et~al.,}{{Mata S{\'a}nchez}
  et~al.}{2023}]{matasanchez2023}
{Mata S{\'a}nchez} D.,  et~al., 2023, \mn@doi [\mnras] {10.1093/mnras/stad203},
  \href {https://ui.adsabs.harvard.edu/abs/2023MNRAS.tmp..227M} {520, 2217}

\bibitem[\protect\citeauthoryear{{Mewe}, {Gronenschild}  \& {van den
  Oord}}{{Mewe} et~al.}{1985}]{mewe1985}
{Mewe} R.,  {Gronenschild} E.~H.~B.~M.,   {van den Oord} G.~H.~J.,  1985,
  \aaps, \href {https://ui.adsabs.harvard.edu/abs/1985A&AS...62..197M} {62,
  197}

\bibitem[\protect\citeauthoryear{{Mewe}, {Lemen}  \& {van den Oord}}{{Mewe}
  et~al.}{1986}]{mewe1986}
{Mewe} R.,  {Lemen} J.~R.,   {van den Oord} G.~H.~J.,  1986, \aaps, \href
  {https://ui.adsabs.harvard.edu/abs/1986A&AS...65..511M} {65, 511}

\bibitem[\protect\citeauthoryear{{Oliveira}, {Rodrigues}, {Cieslinski},
  {Jablonski}, {Silva}, {Almeida}, {Rodr{\'\i}guez-Ardila}  \&
  {Palhares}}{{Oliveira} et~al.}{2017}]{oliveira2017}
{Oliveira} A.~S.,  {Rodrigues} C.~V.,  {Cieslinski} D.,  {Jablonski} F.~J.,
  {Silva} K.~M.~G.,  {Almeida} L.~A.,  {Rodr{\'\i}guez-Ardila} A.,   {Palhares}
  M.~S.,  2017, \mn@doi [\aj] {10.3847/1538-3881/aa610d}, \href
  {https://ui.adsabs.harvard.edu/abs/2017AJ....153..144O} {153, 144}

\bibitem[\protect\citeauthoryear{{Oliveira}, {Rodrigues}, {Martins},
  {Palhares}, {Silva}, {Lima}  \& {Jablonski}}{{Oliveira}
  et~al.}{2020}]{oliveira2020}
{Oliveira} A.~S.,  {Rodrigues} C.~V.,  {Martins} M.,  {Palhares} M.~S.,
  {Silva} K.~M.~G.,  {Lima} I.~J.,   {Jablonski} F.~J.,  2020, \mn@doi [\aj]
  {10.3847/1538-3881/ab6ded}, \href
  {https://ui.adsabs.harvard.edu/abs/2020AJ....159..114O} {159, 114}

\bibitem[\protect\citeauthoryear{{Pogrosheva} et~al.,}{{Pogrosheva}
  et~al.}{2018}]{Pogrosheva2018}
{Pogrosheva} T.,  et~al., 2018, The Astronomer's Telegram, \href
  {https://ui.adsabs.harvard.edu/abs/2018ATel11620....1P} {11620, 1}

\bibitem[\protect\citeauthoryear{{Predehl} et~al.,}{{Predehl}
  et~al.}{2021}]{erosita2021}
{Predehl} P.,  et~al., 2021, \mn@doi [\aap] {10.1051/0004-6361/202039313},
  \href {https://ui.adsabs.harvard.edu/abs/2021A&A...647A...1P} {647, A1}

\bibitem[\protect\citeauthoryear{{Prochaska} et~al.,}{{Prochaska}
  et~al.}{2020a}]{pypeit:zenodo}
{Prochaska} J.~X.,  et~al., 2020a, {pypeit/PypeIt: Release 1.0.0},
  \mn@doi{10.5281/zenodo.3743493}

\bibitem[\protect\citeauthoryear{Prochaska et~al.,}{Prochaska
  et~al.}{2020b}]{pypeit:joss_pub}
Prochaska J.~X.,  et~al., 2020b, \mn@doi [Journal of Open Source Software]
  {10.21105/joss.02308}, 5, 2308

\bibitem[\protect\citeauthoryear{{Romero-Colmenero}, {Potter}, {Buckley},
  {Barrett}  \& {Vrielmann}}{{Romero-Colmenero}
  et~al.}{2003}]{Romero-Colmenero2003}
{Romero-Colmenero} E.,  {Potter} S.~B.,  {Buckley} D.~A.~H.,  {Barrett} P.~E.,
   {Vrielmann} S.,  2003, \mn@doi [\mnras] {10.1046/j.1365-8711.2003.06209.x},
  \href {https://ui.adsabs.harvard.edu/abs/2003MNRAS.339..685R} {339, 685}

\bibitem[\protect\citeauthoryear{{Salvi}, {Ramsay}, {Cropper}, {Buckley}  \&
  {Stobie}}{{Salvi} et~al.}{2002}]{2002MNRAS.331..488S}
{Salvi} N.,  {Ramsay} G.,  {Cropper} M.,  {Buckley} D.~A.~H.,   {Stobie} R.~S.,
   2002, \mn@doi [\mnras] {10.1046/j.1365-8711.2002.05216.x}, \href
  {https://ui.adsabs.harvard.edu/abs/2002MNRAS.331..488S} {331, 488}

\bibitem[\protect\citeauthoryear{{Saxton}, {Read}, {Esquej}, {Freyberg},
  {Altieri}  \& {Bermejo}}{{Saxton} et~al.}{2008}]{xmmsl}
{Saxton} R.~D.,  {Read} A.~M.,  {Esquej} P.,  {Freyberg} M.~J.,  {Altieri} B.,
   {Bermejo} D.,  2008, \mn@doi [\aap] {10.1051/0004-6361:20079193}, \href
  {https://ui.adsabs.harvard.edu/abs/2008A&A...480..611S} {480, 611}

\bibitem[\protect\citeauthoryear{{Scargle}}{{Scargle}}{1982}]{scargle1982}
{Scargle} J.~D.,  1982, \mn@doi [\apj] {10.1086/160554}, \href
  {https://ui.adsabs.harvard.edu/abs/1982ApJ...263..835S} {263, 835}

\bibitem[\protect\citeauthoryear{{Schwope}, {Mantel}  \& {Horne}}{{Schwope}
  et~al.}{1997}]{1997A&A...319..894S}
{Schwope} A.~D.,  {Mantel} K.~H.,   {Horne} K.,  1997, \mn@doi [\aap]
  {10.48550/arXiv.astro-ph/9701094}, \href
  {https://ui.adsabs.harvard.edu/abs/1997A&A...319..894S} {319, 894}

\bibitem[\protect\citeauthoryear{{Schwope}, {Horne}, {Steeghs}  \&
  {Still}}{{Schwope} et~al.}{2011}]{Schwope2011}
{Schwope} A.~D.,  {Horne} K.,  {Steeghs} D.,   {Still} M.,  2011, \mn@doi
  [\aap] {10.1051/0004-6361/201016373}, \href
  {https://ui.adsabs.harvard.edu/abs/2011A&A...531A..34S} {531, A34}

\bibitem[\protect\citeauthoryear{{Schwope}, {Knauff}, {Kurpas}, {Salvato},
  {Stelzer}, {St{\"u}tz}  \& {Tub{\'\i}n-Arenas}}{{Schwope}
  et~al.}{2024}]{schwope2024}
{Schwope} A.~D.,  {Knauff} K.,  {Kurpas} J.,  {Salvato} M.,  {Stelzer} B.,
  {St{\"u}tz} L.,   {Tub{\'\i}n-Arenas} D.,  2024, \mn@doi [\aap]
  {10.1051/0004-6361/202450537}, \href
  {https://ui.adsabs.harvard.edu/abs/2024A&A...690A.243S} {690, A243}

\bibitem[\protect\citeauthoryear{{Shafter}, {Reinsch}, {Beuermann}, {Misselt},
  {Buckley}, {Burwitz}  \& {Schwope}}{{Shafter} et~al.}{1995}]{shafter1995}
{Shafter} A.~W.,  {Reinsch} K.,  {Beuermann} K.,  {Misselt} K.~A.,  {Buckley}
  D.~A.~H.,  {Burwitz} V.,   {Schwope} A.~D.,  1995, \mn@doi [\apj]
  {10.1086/175527}, \href
  {https://ui.adsabs.harvard.edu/abs/1995ApJ...443..319S} {443, 319}

\bibitem[\protect\citeauthoryear{{Shaw} et~al.,}{{Shaw}
  et~al.}{2020}]{2020MNRAS.498.3457S}
{Shaw} A.~W.,  et~al., 2020, \mn@doi [\mnras] {10.1093/mnras/staa2592}, \href
  {https://ui.adsabs.harvard.edu/abs/2020MNRAS.498.3457S} {498, 3457}

\bibitem[\protect\citeauthoryear{{Silber}}{{Silber}}{1992}]{silberphd}
{Silber} A.~D.,  1992, PhD thesis, Massachusetts Institute of Technology

\bibitem[\protect\citeauthoryear{{Singh} et~al.,}{{Singh}
  et~al.}{1995}]{Singh1995}
{Singh} K.~P.,  et~al., 1995, \mn@doi [\apjl] {10.1086/309755}, \href
  {https://ui.adsabs.harvard.edu/abs/1995ApJ...453L..95S} {453, L95}

\bibitem[\protect\citeauthoryear{{Smith} \& {Dhillon}}{{Smith} \&
  {Dhillon}}{1998}]{Smith:1998}
{Smith} D.~A.,  {Dhillon} V.~S.,  1998, \mn@doi [\mnras]
  {10.1046/j.1365-8711.1998.02065.x}, \href
  {https://ui.adsabs.harvard.edu/abs/1998MNRAS.301..767S} {301, 767}

\bibitem[\protect\citeauthoryear{{Spruit}}{{Spruit}}{1998}]{spruit1998}
{Spruit} H.~C.,  1998, \mn@doi [arXiv e-prints]
  {10.48550/arXiv.astro-ph/9806141}, \href
  {https://ui.adsabs.harvard.edu/abs/1998astro.ph..6141S} {pp
  astro--ph/9806141}

\bibitem[\protect\citeauthoryear{{Sunyaev} et~al.,}{{Sunyaev}
  et~al.}{2021}]{Sunyaev2021}
{Sunyaev} R.,  et~al., 2021, \mn@doi [\aap] {10.1051/0004-6361/202141179},
  \href {https://ui.adsabs.harvard.edu/abs/2021A&A...656A.132S} {656, A132}

\bibitem[\protect\citeauthoryear{{Swihart}, {Strader}, {Chomiuk}, {Aydi},
  {Sokolovsky}, {Ray}  \& {Kerr}}{{Swihart} et~al.}{2022}]{swihart2022}
{Swihart} S.~J.,  {Strader} J.,  {Chomiuk} L.,  {Aydi} E.,  {Sokolovsky} K.~V.,
   {Ray} P.~S.,   {Kerr} M.,  2022, \mn@doi [\apj] {10.3847/1538-4357/aca2ac},
  \href {https://ui.adsabs.harvard.edu/abs/2022ApJ...941..199S} {941, 199}

\bibitem[\protect\citeauthoryear{{Szkody}}{{Szkody}}{1998}]{szkody1998}
{Szkody} P.,  1998, in {Howell} S.,  {Kuulkers} E.,   {Woodward} C.,  eds,
  Astronomical Society of the Pacific Conference Series Vol. 137, Wild Stars in
  the Old West. p.~18

\bibitem[\protect\citeauthoryear{{Thomas} \& {Reinsch}}{{Thomas} \&
  {Reinsch}}{1996}]{1996A&A...315L...1T}
{Thomas} H.~C.,  {Reinsch} K.,  1996, \aap, \href
  {https://ui.adsabs.harvard.edu/abs/1996A&A...315L...1T} {315, L1}

\bibitem[\protect\citeauthoryear{{Tovmassian}, {Greiner}, {Zickgraf}, {Kroll},
  {Krautter}, {Thiering}, {Zharykov}  \& {Serrano}}{{Tovmassian}
  et~al.}{1997}]{Tovmassian1997}
{Tovmassian} G.~H.,  {Greiner} J.,  {Zickgraf} F.~J.,  {Kroll} P.,  {Krautter}
  J.,  {Thiering} I.,  {Zharykov} S.~V.,   {Serrano} A.,  1997, \mn@doi [\aap]
  {10.48550/arXiv.astro-ph/9609166}, \href
  {https://ui.adsabs.harvard.edu/abs/1997A&A...328..571T} {328, 571}

\bibitem[\protect\citeauthoryear{{Voikhanskaya}}{{Voikhanskaya}}{1986}]{Voikhanskaya1986}
{Voikhanskaya} N.~F.,  1986, Soviet Astronomy Letters, \href
  {https://ui.adsabs.harvard.edu/abs/1986SvAL...12..196V} {12, 196}

\bibitem[\protect\citeauthoryear{{Wachter}, {Leach}  \& {Kellogg}}{{Wachter}
  et~al.}{1979}]{wstat}
{Wachter} K.,  {Leach} R.,   {Kellogg} E.,  1979, \mn@doi [\apj]
  {10.1086/157084}, \href
  {https://ui.adsabs.harvard.edu/abs/1979ApJ...230..274W} {230, 274}

\bibitem[\protect\citeauthoryear{{Wilms}, {Allen}  \& {McCray}}{{Wilms}
  et~al.}{2000}]{wilms2000}
{Wilms} J.,  {Allen} A.,   {McCray} R.,  2000, \mn@doi [\apj] {10.1086/317016},
  \href {https://ui.adsabs.harvard.edu/abs/2000ApJ...542..914W} {542, 914}

\bibitem[\protect\citeauthoryear{{Wright} et~al.,}{{Wright}
  et~al.}{2019}]{allwise}
{Wright} E.~L.,  et~al., 2019, {AllWISE Source Catalog}, NASA IPAC DataSet,
  IRSA1, \mn@doi{10.26131/IRSA1}

\bibitem[\protect\citeauthoryear{{Zhao}, {Zhao}, {Chu}, {Jing}  \&
  {Deng}}{{Zhao} et~al.}{2012}]{Zhao_2012}
{Zhao} G.,  {Zhao} Y.-H.,  {Chu} Y.-Q.,  {Jing} Y.-P.,   {Deng} L.-C.,  2012,
  \mn@doi [Research in Astronomy and Astrophysics]
  {10.1088/1674-4527/12/7/002}, \href
  {https://ui.adsabs.harvard.edu/abs/2012RAA....12..723Z} {12, 723}

\makeatother
\end{thebibliography}








\bsp	
\label{lastpage}
\end{document}